\documentclass[twocolumn,showpacs,aps,amsmath,amssymb,prl,superscriptaddress]{revtex4-1}   
\usepackage{amsfonts}
\usepackage{bm}
\usepackage{setspace}
\usepackage{graphicx}
\usepackage{color}
\usepackage{multirow}
\usepackage{verbatim}
\usepackage{float}
\usepackage[T1]{fontenc}
\usepackage[normalem]{ulem}
\newcommand{\be}{\begin{equation}}
\newcommand{\ee}{\end{equation}}
\newcommand{\bea}{\begin{eqnarray}}
\newcommand{\eea}{\end{eqnarray}}
\newcommand{\bsube}{\begin{subequations}}
\newcommand{\esube}{\end{subequations}}
\newcommand{\Eq}[1]{Eq.~(\ref{#1})}
\newcommand{\Eqs}[1]{Eqs.~(\ref{#1})}
\newcommand{\Fig}[1]{Fig.~\ref{#1}}

\newcommand{\ua}{\uparrow}
\newcommand{\da}{\downarrow}
\newcommand{\pt}{\partial_t}
\newcommand{\br}{\mathbf{r}}
\newcommand{\bR}{\mathbf{R}}

\newcommand{\no}{\nonumber}

\newcommand{\ep}{\epsilon}
\newcommand{\Ep}{\mathcal E}

\newcommand{\red}[1]{\textcolor[rgb]{1.00,0.00,0.00}{#1}}

\def\bdu#1{\underline{\underline{\bf{#1}}}}

\newcommand{\openbox}{\leavevmode
  \hbox to.77778em{%
  \hfil\vrule
  \vbox to.675em{\hrule width.6em\vfil\hrule}%
  \vrule\hfil}}

\begin{document}

\title{Beyond Born-Oppenheimer Time-Dependent Density Functional Theory}

\author{Chen Li} \email{chenlichem@pku.edu.cn}
\affiliation{Max Planck Institute of Microstructure Physics, Weinberg 2, 06120, Halle, Germany}
\affiliation{Fritz Haber Center for Molecular Dynamics, Institute of Chemistry, The Hebrew University of Jerusalem, Jerusalem 91904 Israel}
\affiliation{Beijing National Laboratory for Molecular Sciences, College of Chemistry and Molecular Engineering, Peking University, Beijing 100871, China}

\author{Ryan Requist}
\affiliation{Max Planck Institute of Microstructure Physics, Weinberg 2, 06120, Halle, Germany}
\affiliation{Fritz Haber Center for Molecular Dynamics, Institute of Chemistry, The Hebrew University of Jerusalem, Jerusalem 91904 Israel}
\affiliation{Division of Theoretical Physics, Ru\dj er Bo\v{s}kovi\'{c} Institute, Zagreb 10000, Croatia}

\author{E. K. U. Gross}
\affiliation{Max Planck Institute of Microstructure Physics, Weinberg 2, 06120, Halle, Germany}
\affiliation{Fritz Haber Center for Molecular Dynamics, Institute of Chemistry, The Hebrew University of Jerusalem, Jerusalem 91904 Israel}

\date{\today}

\begin{abstract}

We formulate a time-dependent density functional theory for 
the coupled dynamics of electrons and nuclei that goes beyond the Born-Oppenheimer (BO) approximation. We prove that the time-dependent marginal nuclear probability density $|\chi({\bdu R},t)|^2$, the conditional electronic density $n_{\bdu R}(\br,t)$, and the current density $\bm J_{\bdu R}(\br,t)$ are sufficient to uniquely determine the full time-evolving electron-nuclear wave function, and thus the  dynamics of all observables. 
Moreover, we propose a time-dependent Kohn-Sham scheme which reproduces the exact conditional electronic density and current density and the exact N-body nuclear density.
The remaining task is to look for functional approximations for the Kohn-Sham exchange-correlation scalar and vector potentials.
Using a model driven proton transfer system, we numerically demonstrate that the adiabatic extension of a beyond-BO ground state functional captures the dominant nonadiabatic effects in the regime of slow driving.

\end{abstract}


\maketitle

%
%
Many fundamental processes in physics, chemistry, and materials science involve the coupled dynamics of electrons and nuclei, representing a challenging quantum many-body problem that requires methods balancing computational efficiency with accuracy.
Within the Born-Oppenheimer (BO) approximation, density functional theory (DFT) has emerged as a highly successful framework for treating the many-electron problem by reformulating the Schr\"odinger equation (SE) in terms of the electron density as reduced fundamental variable. \cite{Hohenberg64B864, Kohn651133} 
Using the standard density functional approximations (DFAs), 
one can reliably obtain an accurate ground-state potential energy surface (PES),  which forms the basis for simulations of quantum or classical nuclear dynamics. Although this approach neglects the back-reaction of nuclear motion on the electronic state, it offers a reasonably accurate description for a wide range of adiabatic chemical processes that do not involve significant electronic excitations out of the ground state.

The coupled motion between electrons and nuclei, however, could induce electronic excitations, whose description requires time-dependent density functional theory (TDDFT). \cite{Runge84997} One common approach is to combine real-time TDDFT with Ehrenfest molecular dynamics, where nuclei evolve on a mean-field PES. \cite{andrade2009, You241100}.
However, in photochemical processes, strong nonadiabatic coupling (NAC) such as that occurring at conical intersections drives significant transitions between adiabatic electronic states with substantially different PES landscapes, leading to dynamics which cannot be captured by a single BO or mean-field PES. A rigorous treatment of the full electron-nuclear wavefunction can be pursued within the Born--Huang (BH) expansion \cite{Born1954}, wherein the nuclear wavefunction evolves over multiple static BO PESs, with population transfer between them mediated by NACs. A standard way to use DFT/TDDFT in this context involves ground state and linear-response calculations of adiabatic PESs and NACs. While the BH framework is well-suited to molecules, whose spectrum is dominated by a finite number of electronic states, it becomes challenging to implement when the spectrum is continuous, as in metals or materials with electrons excited to a conduction band. Examples include electron-phonon energy transfer, as occurs in the dynamics of coherent phonons, \cite{Dekorsy2000, Caruso107054102} chiral phonons, excitons and polarons, \cite{Corby211136} and laser-induced structural phase transitions in solids. \cite{Frigge17207, Nicholson18821}.

In such cases, it is natural to hypothesize that the net effect of occupying a distribution of excited-state PESs can be represented by a single effective time-dependent PES governing nuclear motion. This concept can be formally realized through the exact factorization approach. \cite{hunter75237,gidopoulos2014,abedi10123002, Abedi1222A530}.
In particular, it has been shown that the exact total wave function $\Psi(\bdu r, \bdu R,t)$ can be factorized into a marginal nuclear wave function $\chi(\bdu R,t)$ and a conditional electronic wave function $\Phi_{\bdu R}(\bdu r,t)$; here $\bdu r = (\br_1, \br_2, \cdots, \br_{N_e})$ denotes the electronic coordinates and $\bdu R=(\bR_1, \bR_2, \cdots, \bR_{N_n})$ denotes the nuclear coordinates.
The full TDSE for $\Psi$ is equivalent to the following set of coupled equations for $\chi$ and $\Phi_{\bdu R}$:
{\fontsize{9.}{10}
\begin{align}
    i\pt \Phi_{\bdu R}(\bdu r,t) &= \Big[\hat H_{el}(\bdu R, \bdu r, t) - \ep(\bdu R,t)\Big]\Phi_{\bdu R}(\bdu r,t), \label{Phi} \\
    i\pt \chi(\bdu R,t) &= \Big[\sum_{a=1}^{N_n} \frac{1}{2M_a}\Big(-i\nabla_{\bR_a} + \bm A_a(\bdu R,t)\Big)^2  \nonumber \\
     &\quad+V_{\rm ext}^n(\bdu R,t)+\ep(\bdu R,t)\Big]\chi(\bdu R,t). \label{chi}
\end{align}
}
Here we have set $\hbar=1$ and $V_{\rm ext}^n$ is the external potential acting on the nuclei.
The equation for $\chi$ is a simple TDSE with time-dependent scalar and vector potentials $(\Ep, \bm A_l)$ that are unique up to a gauge choice; this is very different from the BH formalism where the nuclear wave amplitude evolves on all static BO PESs and the population on different surfaces is exchanged through the nonadiabatic couplings. The equation for $\Phi_{\bdu R}$ is reminiscent of the electron-only problem at each nuclear configuration $\bdu R$ but with $\hat H_{el}(\bdu R, \bdu r, t) = \hat H_{\rm BO}+ \hat U_{en}$ containing both the BO Hamiltonian and a $\nabla_{\mathbf{R}_l}$-gradient correction $\hat U_{en}$ (see \cite{abedi10123002, Abedi1222A530} for the formula).
\Eqs{Phi}--\eqref{chi} provide an ideal starting point for 
density functionalization. In particular, through replacing $\Phi_{\bdu R}$ by some electronic densities and reformulating \Eq{Phi} as a time-dependent conditional Kohn-Sham scheme,
we aim at reproducing the exact time-evolving densities. 

Given the inherently conditional nature of this quantity, a natural choice is to introduce $n_{\bdu R}(\br, t)$, the conditional electronic density, as the fundamental variable. 
Besides $n_{\bdu R}(\br, t)$, we note that there is another quantity, the conditional electronic current density $\bm J_{\bdu R}(\br,t)$, which deserves consideration as an additional fundamental variable. This is because electrons typically follow nuclear motion in a coupled electron-nuclear dynamics, yielding a nontrivial electronic current. \cite{Schild163316} Formulating the theory with $\bm J_{\bdu R}(\br,t)$ extends time-dependent current density functional theory (TDCDFT) beyond purely electronic systems, which is the approach adopted in this Letter.  

This formulation involves reducing the fundamental variable from the full wavefunction $\Psi$ to the set $(|\chi|^2, n_{\bdu R}, \bm J_{\bdu R})$, a reduction that entails a significant loss of information. To justify this choice, it is essential to establish an extension of the Vignale theorem \cite{Vignale04201102} for the coupled electron-nuclear problem.
The key lies in establishing a one-to-one mapping between $(n_{\bdu R}, \bm J_{\bdu R})$ with some scalar and vector potential, which is not trivial for a two-component system.
Instead of using conditional densities, we show that a better starting point is using joint electronic densities and current densities, defined by 
\begin{align}
	\rho(\br, \bdu R, t) &= N_e \int |\Psi(\bdu r, \bdu R, t)|^2 d\br_2 \cdots d\br_{N_e}, \\
	\bm J(\br, \bdu R, t) &= \frac{-iN_e}{2m_e} \int \Big(\Psi^*\nabla_{\br}\Psi - \Psi\nabla_{\br}\Psi^*\Big) d\br_2 \cdots d\br_{N_e},
\end{align}
which are 1-body quantities in electrons and N-body in nuclei.
One can verify that our fundamental variables are derivable from $\rho$ and $\bm J$ through the following relations: $|\chi(\bdu R, t)|^2 = \frac{1}{N_e}\int \rho(\br, \bdu R, t)d\br$, $n_{\bdu R}(\br, t)=\rho(\br, \bdu R, t)/|\chi(\bdu R, t)|^2$ and $\bm J_{\bdu R}(\br, t)=\bm J(\br, \bdu R, t)/|\chi(\bdu R, t)|^2$. Thus, the joint densities carry exactly the same information as \{$|\chi|^2$, $n_{\bdu R}$, $\bm J_{\bdu R}$\}. Using $\rho$ and $\bm J$, we state our theorem regarding a density-to-potential mapping as follows.

\emph{Theorem.}
Consider an electron-nuclear system described by the time-dependent Hamiltonian
\fontsize{9}{10}{
\begin{align}
	\hat H &= -\sum_{a}\frac{1}{2M_a}\nabla_{\bR_a}^2 + \sum_{k}\frac{1}{2m_e}\Big(\hat {\bm p}_k + \bm A(\br, \bdu R,t) \Big)^2 \nonumber \\
	&\quad  + \sum_{a<b} U(R_{ab}) + \sum_{j<k} W(r_{jk}) + \sum_{k} V(\br_k, \bdu R, t).
\end{align}
}
Here $\hat {\bm p}_k = -i\nabla_{\br_k}$ is the canonical momentum operator of electron $k$, $R_{ab}= |\bR_a - \bR_b|$ and $r_{jk} =|\br_j - \br_k|$. $V(\br, \bdu R, t)$ and $\bm A(\br, \bdu R, t)$ are the time-dependent electron-nuclear scalar and vector interaction potentials, respectively, and assumed to be analytic functions of $t$ at $t=0$. We show that under reasonable assumptions, i.e., the Taylor series expansion of the potentials have finite radius of convergence, the joint 1-electron N-nucleus density $\rho(\br, \bdu R, t)$ and the current density $\bm J(\br, \bdu R, t)$ can be obtained under the influence of $\hat H$ from a given initial state $\Psi(0)$ and can also be obtained under the influence of $\hat H'$,
{\fontsize{9.}{10}
\begin{align}
	\hat H' &= -\sum_{a}\frac{1}{2M_a}\nabla_{\bR_a}^2  + \sum_{k}\frac{1}{2m_e}\Big(\hat {\bm p}_k + \bm A'(\br, \bdu R,t) \Big)^2 \no \\
	&\quad  + \sum_{a<b} U(R_{ab}) + \sum_{j<k} W'(r_{jk}) + \sum_{k} V'(\br_k, \bdu R, t),
\end{align}
}
starting from an initial state $\Psi'(0)$ that gives the same $\rho$ and $\bm J$ as $\Psi(0)$ at $t=0$. The potentials $V'(\br, \bdu R, t)$ and $\bm A'(\br, \bdu R,t)$ are uniquely determined by $V(\br, \bdu R, t)$ and $\bm A(\br, \bdu R,t)$, $\Psi(0)$ and $\Psi'(0)$, up to gauge transformations of the form 
\begin{align}
	\tilde V(\br, \bdu R, t) &\rightarrow V'(\br, \bdu R, t) - \pt \Lambda(\br, \bdu R, t),  \no \\
	\bm \tilde A(\br, \bdu R, t) &\rightarrow \bm A'(\br, \bdu R, t) + \nabla_{\br} \Lambda(\br, \bdu R, t).
\end{align}

\emph{Proof.} Given potentials $V(\br, \bdu R, t)$ and $\bm A(\br, \bdu R, t)$, one can always make a gauge transformation such that 
the scalar potential vanishes. This can be achieved by setting 
\begin{align}
	\pt \Lambda(\br, \bdu R, t) = V(\br, \bdu R, t)
\end{align}
with initial condition $\Lambda(\br, \bdu R, 0) = 0$. The same argument also applies to the primed potentials. Therefore, for the convenience of derivation, let us assume $V$ and $V'$ have been gauged away. Then it suffices to construct the differential equation that uniquely determines $\bm A'(\br, \bdu R, t)$. 

We begin by denoting $\bm {\hat v}_k(t) = \frac{1}{m_e}\Big(\bm {\hat p}_k+\bm A(\br, \bdu R, t)\Big)$ and rewriting $\rho(\br, \bdu R, t) = \langle \Psi |\hat n(\br)|\Psi\rangle_{\bdu r}$ and $\bm J(\br, \bdu R, t) = \langle \Psi |\hat {\bm j}(\br,t)|\Psi\rangle_{\bdu r}$, where $\hat n(\br) = \sum_{k}\delta(\br-\br_k)$ and $\bm {\hat j}(\br, t) = \frac{1}{2}\sum_{k}\{\bm {\hat v}_k(t), \delta(\br-\br_k) \}$ are electron density and current density operators, respectively. Here $\{\hat A, \hat B\} = \hat A \hat B+ \hat B\hat A$ is the anticommutator. Using Heisenberg equation of motion, we can deduce the time derivative of the current density,
\begin{align}
	\pt \bm J &= \langle \Psi |i[\hat H, \bm {\hat j}(\br, t)]|\Psi\rangle_{\bdu r} + \langle \Psi |\pt\bm {\hat j}(\br, t)|\Psi\rangle_{\bdu r} . \label{ptJe}
\end{align}
By straightforward algebra, one can show $\langle \Psi |\pt\bm {\hat j}(\br, t)|\Psi\rangle_{\bdu r} = \frac{1}{m_e}\rho(\br, \bdu R, t)\pt \bm A$.
To evaluate the first term on the right hand side (RHS) of \Eq{ptJe}, we decompose $\hat H = \hat H_n + \hat H_{\rm BO}$, where $\hat H_n$ is the nuclear kinetic energy operator. Invoking Vignale's result in Ref \cite{Vignale04201102}, we arrive at (details can be found in the supplemental information \cite{supp})
\begin{align}
	  & \langle \Psi |i[\hat H_{\rm BO}, \bm {\hat j}(\br, t)]|\Psi\rangle_{\bdu r} \no\\
	=&  \frac{1}{m_e}\Big[-\bm J\times (\nabla_{\br} \times \bm A) 
	+ \bm F(\br, \bdu R, t)\Big]+ \nabla_{\br} \cdot \bm \sigma(\br, \bdu R, t).
\end{align}
Here
\begin{align}
	\bm F(\br, \bdu R, t) = -\langle \Psi | \sum_k \delta(\br-\br_k)\sum_{l\neq k} \nabla_{\br_k} W(|\br_k-\br_l|)|\Psi\rangle_{\bdu r},
\end{align}
and
\begin{align}
	\sigma_{\alpha \beta}(\br, \bdu R, t) = -\frac{1}{4}\langle \Psi | \sum_k \{\hat v_k^\beta, \{\hat v_k^\alpha, \delta(\br-\br_k)\}\} |\Psi\rangle_{\bdu r}
\end{align}
is a stress tensor, with $\alpha$ and $\beta$ referring to the Cartesian indices. Denoting $\bm S = \langle \Psi |i[\hat H_n, \bm {\hat j}(\br, t)]|\Psi\rangle_{\bdu r}$, we have
\begin{equation}
	\fontsize{10.}{10} \pt \bm J = \frac{1}{m_e}\Big[\rho\pt \bm A- \bm J\times (\nabla_{\br} \times \bm A) + \bm F\Big] 
	+ \nabla_{\br} \cdot \bm \sigma + \bm S. 
\end{equation}
Since $\pt \bm J$ is identical for the unprimed and primed system, it follows
\begin{align}
	&\quad \frac{1}{m_e}\Big[\rho\pt \Delta \bm A- \bm J\times (\nabla_{\br} \times \Delta \bm A) \Big] + \Delta \bm Q =0. \label{DeltaA}
\end{align}
Here $\Delta \bm A = \bm A' - \bm A$ and $\Delta \bm Q = \bm Q' - \bm Q$, with
\begin{equation}
	\fontsize{10}{10} \bm Q(\br, \bdu R, t) = \frac{\bm F(\br, \bdu R, t)}{m_e}+ \nabla_{\br} \cdot \bm \sigma(\br, \bdu R, t) + \bm S(\br, \bdu R, t), 
\end{equation}
and $\bm Q'$ being the counterpart of the primed system.

At $t=0$, using the equality of the current densities, i.e., $\langle \Psi_0 | \hat {\bm j}(\br, t) |\Psi'_0 \rangle_{\bdu r}=\langle \Psi_0 | \hat {\bm j}(\br, t) |\Psi'_0 \rangle_{\bdu r}$, we can deduce
\begin{align} 
	\rho(\br, \bdu R,0)\Delta \bm A(\br, \bdu R, 0) = \langle \Psi_0 | \bm {\hat j}_p(\br)  |\Psi_0 \rangle_{\bdu r} - \langle \Psi_0' | \bm {\hat j}_p(\br)  |\Psi_0' \rangle_{\bdu r}. \label{DeltaA0}
\end{align}
Here $\bm {\hat j}_p(\br) = \sum_k \{\bm {\hat p}_k, \delta(\br-\br_k)\}$ is the paramagnetic current density operator. \Eq{DeltaA} along with \Eq{DeltaA0} define a partial differential equation for $\Delta \bm A(\br, \bdu R,t)$. By our assumption, $\Delta \bm A$ is Taylor expandable with respect to $t$ with nonzero radius of convergence. Then following Vignale's constructive proof in \cite{Vignale04201102}, we can plug the Taylor series formula $\Delta \bm A(\br, \bdu R,t) = \sum_{k}\Delta \bm A_k(\br, \bdu R) t^k $ into \Eq{DeltaA}, which leads to a recursive relation for $\Delta \bm A_k$. \cite{supp} By \Eq{DeltaA0}, $\Delta \bm A_0$ is known. Therefore, the recursive relation shall uniquely determine all the $\Delta \bm A_k$, and hence $\Delta \bm A$ and $\bm A'$. This completes both the existence and uniqueness proof of our theorem.

By our theorem, the mapping between the potentials $\Big(V(\br, \bdu R, t), \bm A(\br, \bdu R, t)\Big)$ and the densities $\Big(|\chi(\bdu R ,t)|^2, n_{\bdu R}(\br, t), \bm J_{\bdu R}(\br,t)\Big)$ is invertible up to a gauge transformation, allowing us to use these densities as the fundamental variables to reproduce the exact dynamics. 
Additionally, by choosing $W'=0$ we can introduce an auxiliary Kohn-Sham system with scalar and vector potentials $V'$ and $\bm A'$ that reproduce the actual densities. Similar idea has been exploited for the ground state problem in \cite{Fromager24025002}.
Applying exact factorization to the full wave function for this Kohn-Sham system, i.e. $\Psi'(\bdu r, \bdu R, t) = \chi'(\bdu R, t) \Phi'_{\bdu R}(\bdu r,t)$, leads to coupled equations for $\Phi'_{\bdu R}$ and $\chi'$ analogous to \Eqs{Phi}--\eqref{chi} with unprimed quantities replaced by primed ones everywhere. In particular, $\hat H_{el}'(\bdu R, \bdu r, t) = \hat H_{\rm TDKS}^{\rm BO} + \hat U_{en}$, where
\begin{align}
	\hat H_{\rm TDKS}^{\rm BO} &= \sum_{k}\frac{1}{2m_e}\Big(\hat {\bm p}_k + \bm A'(\br, \bdu R,t) \Big)^2 + \sum_{a<b} U(R_{ab})\no \\
	&\quad   + \sum_{k} V'(\br_k, \bdu R, t), \\
	\hat U_{en}[\Phi', \chi'] &= \sum_a \frac{1}{M_a}\Big[\frac{(-i\nabla_{\bR_a}-\bm A_a')^2}{2}\nonumber \\
	&\quad+(\frac{-i\nabla_{\bR_a} \chi'}{\chi'}+\bm A_a')(-\nabla_{\bR_a} -\bm A_a')\Big].
\end{align}
The Hamiltonian governing the nuclear dynamics reads
\begin{align}
	\hat H_n'(\bdu R,t) &= \sum_a \frac{1}{2M_a}\Big(-i\nabla_{\bR_a} + \bm A_a'(\bdu R,t)\Big)^2+\ep'(\bdu R,t),
\end{align} 
with $\bm A_a'(\bdu R,t) = \langle \Phi' | -i\nabla_{\bR_a} |\Phi' \rangle_{\bdu r}$, $\ep'(\bdu R,t) = \langle \Phi' | \hat H_{el}'(\bdu R, \bdu r, t) -i\pt |\Phi' \rangle_{\bdu r}$. 

The equation for $\Phi'_{\bdu R}(\bdu r,t)$ has two unusual, but highly desirable, properties: (i) Owing to its $\nabla_{\bR_l}$ dependence, $\hat U_{en}$ acts like a non-Hermitian operator in the electronic Hilbert space. Consequently, the resulting time-propagation of $\Phi'_{\bdu R}(\bdu r,t)$ is non-unitary despite being norm-conserving by construction. We emphasize that it is precisely this non-unitary evolution that enables the description of electronic decoherence within a single-trajectory approach. \cite{Tu25043075} (ii) Even if the initial state $\Phi'_{\bdu R}(\bdu r,t_0)$ is a determinant of single-particle orbitals, the $\nabla_{\bR_l}^2$ term in $\hat U_{en}$ creates correlations among the electrons, preventing $\Phi'_{\bdu R}(\bdu r,t)$ from staying a single Slater determinant. In fact, a similar feature is known in the traditional Lindblad approach where an initially non-interacting system becomes correlated through an effective, environment-mediated interaction between the particles of the system. \cite{Ruan182490}

For real calculations, because $\hat U_{en}'$ is inversely proportional to the nuclear mass, it can be treated perturbatively. \cite{Schild163316, Tu25043075, Tu2025} Alternatively, in the spirit of traditional KS TDDFT, we adopt, as a working hypothesis, the assumption that the nuclear observable quantities, namely, the N-body density $|\chi(\bdu R,t)|^2$ and current density $\bm J_n({\bdu R},t)$ of the exact dynamics can be reproduced from the time evolution of the marginal nuclear wave amplitude $\chi(\bdu R, t)$ according to \Eq{chi}, and the assumption that the electronic observable quantities, specifically $n_{\bdu R}(\br,t)$ and $\bm J_{\bdu R}(\br,t)$, can be reproduced from the evolution of conditional electronic Kohn-Sham orbitals $\varphi^k_{\bdu R}(\br, t)$ that satisfy the following TDKS equations:
\begin{equation}
	\fontsize{10.}{10} i\pt \varphi_k = \frac{1}{2m_e}\Big(-i\nabla + \bm A_s(\br,\bdu R,t)\Big)^2\varphi_k + v_s(\br,\bdu R,t)\varphi_k. \label{varphi}
 \end{equation} 
Here $v_s(\br,\bdu R,t)$ and $\bm A_s(\br,\bdu R,t)$ are KS scalar and vector potentials, respectively, which are functionals of the fundamental density variables. 
Similar representability assumptions have also been made in the real time propagation of correlated electron-nuclear dynamics.  \cite{Han232186}.

The remaining task is to look for good functionals for $v_s$ and $\bm A_s$ and for the potentials $\ep(\bdu R,t)$ and $\bm A_l(\bdu R, t)$ in \Eq{chi}. Let us start with a simple case where the vector potentials of both $\bm A_l$ and $\bm A_s$ can be gauged away. Then it suffices to consider only the scalar potentials, $\ep$ and $v_{s,\bdu R}$, as functionals of $n_{\bdu R}$ and $|\chi|^2$.
In Ref \cite{Li18084110, Wang25234104}, we have shown that the ground state can be obtained by solving coupled static Kohn-Sham equations and nuclear Schr{\"o}dinger equation. Moreover, we have demonstrated using a model charge transfer system that the major beyond-BO effect (due to the finiteness of the nuclear mass) can be captured by geometric corrections, $v_{\rm geo}$ and $\ep_{\rm geo}$, to the conventional KS potential and BO PES, respectively.

For the time-dependent problem, it is natural to consider the TD extension of these corrections. However, \Eq{chi} shows that the TD PES is defined by $\ep(\bdu R, t) = \langle \Phi_{\bdu R}|\hat H_{el}|\Phi_{\bdu R}\rangle +\langle \Phi_{\bdu R}|-i\pt \Phi_{\bdu R}\rangle $, where we emphasize that $\epsilon_{\rm dyn}\equiv \langle \Phi_{\bdu R}|-i\pt \Phi_{\bdu R}\rangle $ is an extra dynamical contribution to the PES not present in the static case. \cite{Abedi13263001, Agostini133625} 
Nevertheless, as a first approximation we neglect $\epsilon_{\rm dyn}$, which has been shown to be negligible in the adiabatic limit. \cite{supp} Following the strategy of the electron-only TDDFT, we approximate $\ep(\bdu R, t)$ and $v_{s,\bdu R}$ using the adiabatic extension of the ground state functionals. In particular, we can approximate the PES as $\ep=\epsilon_{\rm BO}+\epsilon_{\rm geo}$ with $\epsilon_{\rm BO}$ being our choice of DFA under the BO approximation and $\epsilon_{\rm geo}$ treated by our recently developed local conditional density approximation (LCDA), $\epsilon_{\rm geo}=\epsilon_{\rm geo}[n,\nabla_{\bR_l} n]$; $v_{s,\bdu R}=V(\br, \bdu R, t)+v_{xc, \bdu R}^{\rm BO}(\br,t)+v_{\rm geo,\bdu R}(\br,t)$, with 
$v_{\rm geo}$ derived from the functional derivative of $\epsilon_{\rm geo}$.

\begin{figure}[t]
\includegraphics[width=1.0\columnwidth]{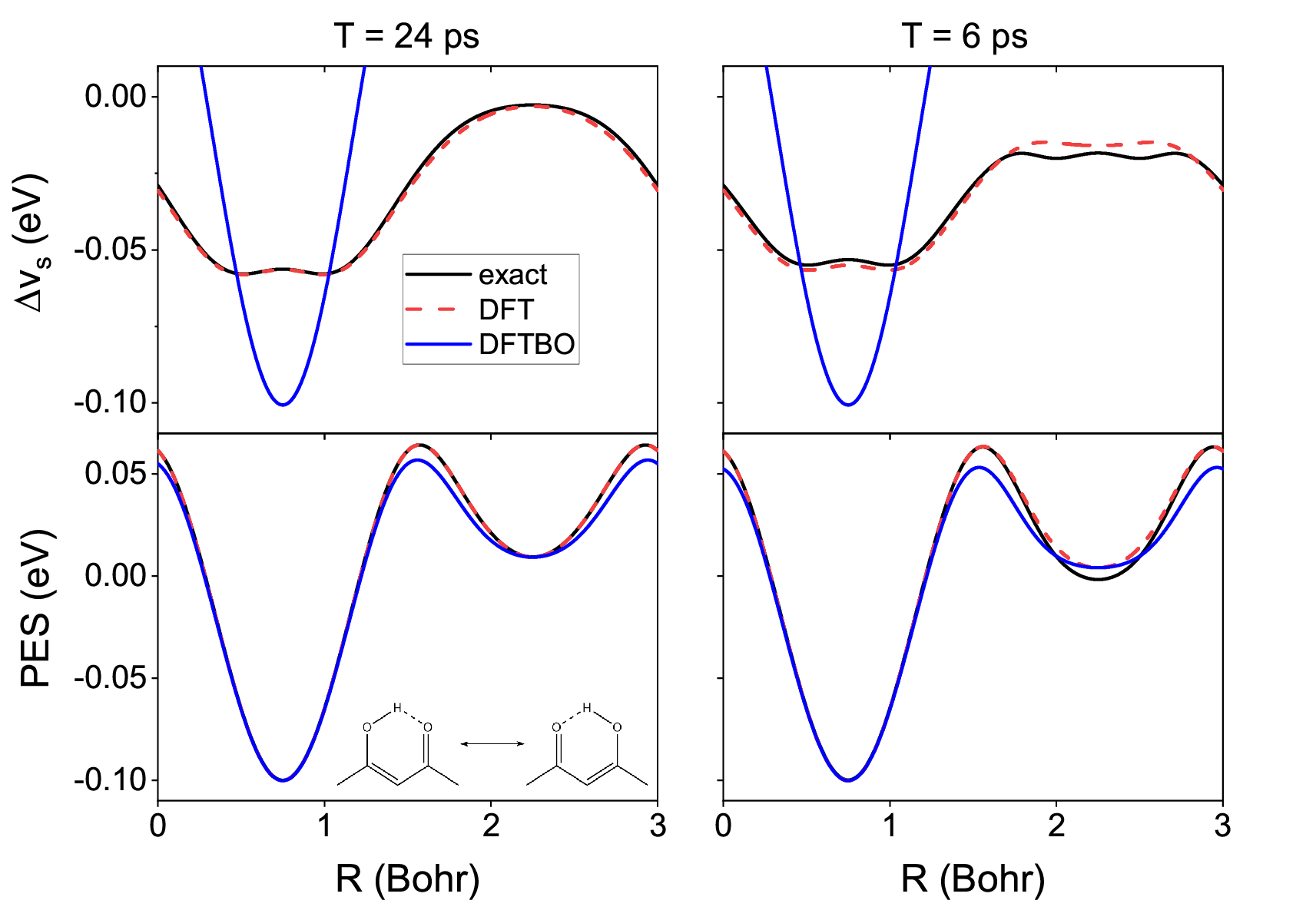}
\centering
\caption{
Comparison between the exact reverse-engineered Kohn-Sham potential $\Delta v_s$ and PES (black solid) with the ones obtained by plugging the exact $n_R$ and $|\chi|$ into the BO (blue solid) and beyond-BO (red dashed) ground state functional in the adiabatic regime ($T=24$ ps) and an intermediate regime ($T=6$ ps). All variables are evaluated at $t=\frac{T}{2}$. The inset illustrates the transformation of two enol tautomers of acetylacetone that is the target of our model. \label{comp_adiabatic}}
\end{figure}

In the following, we apply our beyond-BO TDDFT formalism to a model driven proton transfer process, demonstrating that $\epsilon$ and $v_{s,\bdu R}$ can be accurately approximated by the adiabatic extension of a ground state functional. In order to compare with an exact solution which is unavailable in more realistic systems, we restrict the nuclear configuration space to one dimension and consider a double well model, mimicking a hydrogen transfer reaction through the tautomerism of two enol structures of acetylacetone driven by a bias potential (see \Fig{comp_adiabatic}). Furthermore, we effectively truncate the electronic Hilbert space by using an $R$-dependent two-site Hubbard model as in \cite{Li18084110}. Extension to continuous density has been achieved in \cite{Wang25234104}.
In the basis of the three singlet states, namely, $\varphi_1 = |1_\ua 1_\da\rangle$, $\varphi_2 = \frac{1}{\sqrt 2}(|1_\ua 2_\da\rangle-|1_\da 2_\ua\rangle$), and $\varphi_3 = |2_\ua 2_\da\rangle$, the electronic Hamiltonian is
\begin{equation}
    \bm H_e(R) = \begin{pmatrix}
    U_1 + \Delta \epsilon(R), & -\sqrt 2 \tau(R), & 0 \\
    -\sqrt 2\tau(R), & 0, & -\sqrt 2\tau(R) \\
    0, & -\sqrt 2\tau(R), & U_2 - \Delta \epsilon(R)
    \end{pmatrix}.
\end{equation}
Here $U_i$ are the on-site Hubbard parameters; $\tau(R)$ is the electron hopping energy; and $\Delta \epsilon(R)$ is the on-site energy difference. The electron density operator is defined as $\hat n = \rm{diag}(-1,0,1)$. Moreover, to avoid possible numerical difficulties at the boundary during the time propagation, we choose periodic functions for $\tau(R)$ and $\Delta \epsilon(R)$ with period $L=3$ Bohr. We choose the nuclear mass $M = 2000m_e$ to be about the hydrogen mass, and choose the parameters in the Hubbard model such that the barrier in the ground state PES roughly corresponds to a weak hydrogen bonding energy.
To describe the external driving field, which effectively simulates a nearby polar solvent molecule, \cite{Hanna05244505} we introduce the following time-dependent driving potential,
\begin{equation}
    \bm V(R,t) = V_0 \sin(\frac{2\pi R}{L})\hat n \cos \omega t,
\end{equation}
which couples to both the electrons and nuclei. Here $V_0$ is the amplitude and $\omega \equiv \frac{2\pi}{T}$ is the frequency of the driving potential. We adopt a large $V_0$ to amplify the driving in order to have a stringent test of our DFT functional.
The total time-dependent electron-nuclear Hamiltonian is then given by $\bm {\hat H} = -\frac{1}{2M}\nabla^2 + \bm H_e(R)+\bm V(R,t)$.
In the absence of the driving potential $\bm V$, the ground state PES is a symmetric double well; see the supplemental material \cite{supp} for details of the model and a graphical illustration. 
7
With an external bias potential 
at $t=0$, it lowers the level of the right well (located at $R>
\frac{L}{2}$) and raises the level of the left one ($R<\frac{L}{2}$), so that the instantaneous ground state of the Hamiltonian gives a proton density that mainly populates the right well. We start with such a state as the initial condition and slowly drive the potential for half a time period, by which time the relative energy between the wells is reversed and the proton should transfer to the left well; the electron density changes correspondingly. By performing the time evolution of the exact TDSE, one can compute the exact $n_R$ and $|\chi|$ and compare them with the result of evolving the TDDFT equations of \Eq{varphi} and \Eq{chi}.
However, since the density behavior can be derived from the potentials, which are readily available and less sensitive to error propagation, here we take the exact time-dependent $n_R$ and $|\chi|$ and reverse engineer the corresponding time-evolving KS potential (assuming non-interacting $v$-representability) and PES, and compare them with the adiabatic extension DFT counterparts using the exact $n_R$ and $|\chi|$ as input. This is shown in Figure~\ref{comp_adiabatic}. As we slow down the driving frequency (increase $T$), both the KS potential and the PES from our DFT functional approximations reach better agreement with the exact potentials. For $T\gtrsim 20$ ps, the DFT results essentially overlap with the exact ones; the remaining difference is essentially inherited from errors in the static ground state functional due to the use of an approximate BO functional and the LCDA.
These results validate the adiabatic extension approximation in the adiabatic regime.
Here we also present the results of the BO approximation without the geometric correction, named DFTBO in the figure. Although the PES is well reproduced by DFTBO, we note that this is due to our use of the exact density as input and that the KS potential is drastically wrong. In an actual time evolution, the large errors in the KS potential would lead to a completely incorrect density, which would then feed back into the PES, affecting the nuclear wave function. This comparison between DFT and DFTBO thus indicates the vital role of the geometric correction in the adiabatic regime.

In the diabatic regime ($\omega\rightarrow \infty, T\rightarrow 0$) and some of the intermediate regimes, however, the adiabatic extension approximation should completely fail, as in the electron-only TDDFT. Besides the lack of memory dependence in the adiabatic functional, we note that in the electron-nuclear context, there is an additional missing term, i.e., the $\ep_{\rm dyn}$ term, which plays an important role. \cite{supp} To accurately describe the nonadiabatic regimes, a decent approximation to $\ep_{\rm dyn}$ is needed.

As an additional remark, in our test example we have assumed the non-interacting $v$-representability, which is not generally true for arbitrary regime. In fact, we have found counterexamples in our model for some intermediate regimes (0.02 ps $<T<$ 5 ps). \cite{supp} 
This result is consistent with our expectations, as the assumption in \Eq{varphi} is expected to be valid under perturbative conditions while likely breaking down in the diabatic regime. 
Nevertheless, by incorporating a vector potential in the TDKS equation, this problem is likely to be alleviated. 
This can be better verified by testing on a continuous density model where one can include the electron current density as the fundamental variable. We leave this for future investigation. 

In conclusion, we have formulated a beyond-BO TDDFT that fully incorporates the nuclei.
Here our main focus has been to density functionalize the electronic equations. In practice, one can also apply techniques to simplify the nuclear equations, such as using trajectory-based methods for reconstructing $|\chi(\bdu R)|$ \cite{Martinez96252, Coker95496, Nielsen006543, Shalashilin09244101, Abedi1433001, Agostini14214101, Min15073001, Min173048, Gossel184513}.
Regarding functional approximations, in this Letter, by applying the LCDA functional to a model driven proton transfer system, we have demonstrated the validity of our functional in the adiabatic limit.
As indicated in Ref.~\cite{Li18084110, Wang25234104}, for real systems with continuous densities, the LCDA reduces the search for beyond-BO functionals to the search for a local function of $n_{\bdu R}$. This key simplification opens the door to TDDFT molecular dynamics simulations including nonadiabatic and nuclear quantum effects.

%
%
We thank Dr. Graeme H. Gossel for helpful discussions on time propagation schemes.
Chen Li appreciates funding support from the National Key Research and Development Program of China (2023YFA1507000), the National Science Foundation of China (8200906190) and the Li Ge-Zhao Ning Youth Research Foundation (LGZNQN202403). This project has received funding from the European Research Council (ERC) under the European Union's Horizon 2020 research and innovation programme (grant agreement No ERC-2027-AdG-788890). E. K. U. Gross acknowledges support as Mercator Fellow at the University Duisburg-Essen within SFB 1242 funded by the Deutsche Forschungsgemeinschaft (DFG) Project No 278162697. 

%

\end{document}


\draft

Supporting Information of
%
\vspace{0.1cm}

\begin{center}
{\large \textbf{Beyond Born-Oppenheimer Time-Dependent Density Functional Theory}}
\vspace{0.2cm}

Chen Li$^1$$^,$$^2$$^,$$^3$, Ryan Requist$^1$$^,$$^2$$^,$$^4$ and E. K. U. Gross$^1$$^,$$^2$

{\small
\emph{$^1$Max Planck Institute of Microstructure Physics, Weinberg 2, 06120, Halle, Germany
\\}
\emph{$^2$Fritz Haber Center for Molecular Dynamics, Institute of Chemistry, The Hebrew University of Jerusalem, Jerusalem 91904 Israel
\\}
\emph{$^3$Beijing National Laboratory for Molecular Sciences, College of Chemistry and Molecular Engineering, Peking University, Beijing 100871, China \\}
\emph{$^4$Division of Theoretical Physics, Ru\dj er Bo\v{s}kovi\'{c} Institute, Zagreb 10000, Croatia \\}
}

\vspace{0.5cm}

(Dated: \today)

\end{center}

\linespread{1.0}


\tableofcontents

\clearpage

\renewcommand{\theequation}{S\arabic{equation}}
\renewcommand{\thetable}{S\arabic{table}}
\renewcommand{\thefigure}{S\arabic{figure}}

\section{Some details in the derivation of $\pt \bm J$}

For an electron-only problem $i\pt \Phi = \hat H_{\rm BO}\Phi$ with
\begin{align}
	\hat H_{\rm BO}(t) = \sum_k \Big[\frac{1}{2m_e}[\bm {\hat p}_k^2 + \bm A(\br_k,t)]^2 + V(\br_k, t)\Big] + \sum_{k<l} W(|\br_k-\br_l|)+ \sum_{a<b} U(|\bR_a-\bR_b|),
\end{align}
Here we have suppressed $\bdu R$-dependence of $V, \bm A$ and $\hat H_{\rm BO}$ for brevity.
By the Heisenberg equation of motion, 
\begin{align}
	\pt \bm j(\br,t) = \pt \langle \Phi | \hat {\bm j}(\br,t)|\Phi\rangle_{\bdu r} = i\langle \Phi | [\hat H_{\rm BO}, \hat {\bm j}(\br,t)]|\Phi\rangle_{\bdu r} + \langle \Phi | \pt\hat {\bm j}(\br,t)|\Phi\rangle_{\bdu r}.
\end{align}
Vignale has shown \cite{Vignale04201102} that 
\begin{align}
	&\quad  i\langle \Phi | [\hat H_{\rm BO}, \hat {\bm j}(\br,t)]|\Phi\rangle_{\bdu r} = \frac{1}{m_e}\Big[-\bm j\times (\nabla\times \bm A) + \bm F\Big]+ \nabla \cdot \bm \sigma, \label{commutator}
\end{align}
where 
\begin{align}
	\bm F = -\langle \Phi |\sum_{k}\delta(\br-\br_k)\sum_{l\neq k}\nabla_l W(|\br_k-\br_l|)|\Phi\rangle_{\bdu r},
\end{align}
and 
\begin{align}
	\sigma_{\alpha \beta}(\br, t) = -\frac{1}{4}\langle \Phi |\sum_k\{ {\hat v}_k^{\beta} , \{{\hat v}^\alpha_k, \delta(\br-\br_k)\}\}|\Phi\rangle_{\bdu r}
\end{align}
is a symmetric stress tensor. Here $\bm{\hat v}_k = \frac{1}{m_e}(-i\nabla_{\br_k} + \bm A)$ is the velocity operator and ${\hat v}^\alpha_k$ is one of its Cartesian components.

Now in the electron-nuclear problem, as an intermediate step in the calculation of $\pt J(\br, \bdu R,t)$, we need to evaluate 
$i\langle \Psi | [\hat H_{\rm BO}, \hat {\bm j}(\br,t)]|\Psi\rangle_{\bdu r}$. Using the exact factorization, $\Psi(\bdu r,\bdu R, t) = \chi(\bdu R,t)\Phi_{\bdu R}(\bdu r,t)$. Because the operator $[\hat H_{\rm BO}, \hat {\bm j}(\br,t)]$ does not act on $\chi$, we have
\begin{align}
	i\langle \Psi | [\hat H_{\rm BO}, \hat {\bm j}(\br,t)]|\Psi\rangle_{\bdu r} = i|\chi|^2 \langle \Phi | [\hat H_{\rm BO}, \hat {\bm j}(\br,t)]|\Phi\rangle_{\bdu r}.
\end{align}
By \Eq{commutator}, we only need to replace the terms on its right hand side (RHS) by the corresponding conditional quantities. Note that $|\chi|^2 \bm J_{\bdu R}(\br, t) = \bm J(\br, \bdu R, t)$ gives the joint current density. Moreover, when $\bm F$ and $\bm \sigma$ are multiplied by $|\chi|^2$, one can simply absorb $|\chi|^2$ into the bra and ket through replacing $\Phi$ by $\Psi$ (let us rename the formulas as $\tilde {\bm F}$ and $\tilde {\bm \sigma}$). Thus, one ultimately obtains
\begin{align}
	i\langle \Psi | [\hat H_{\rm BO}, \hat {\bm j}(\br,t)]|\Psi\rangle_{\bdu r} = \frac{1}{m_e}\Big[-\bm J(\br, \bdu R, t)\times (\nabla\times \bm A) + \tilde {\bm F}\Big]+ \nabla \cdot  \tilde {\bm \sigma}.
\end{align}

\section{Some details of the existence proof of our theorem}

In the main text we have derived the partial differential equation for $\Delta \bm A(\br, \bdu R, t)$ as the following:
\begin{align}
	&\quad \frac{1}{m_e}\Big[\rho\pt \Delta \bm A- \bm J\times (\nabla_{\br} \times \Delta \bm A) \Big] + \Delta \bm Q =0, \label{DeltaA}
\end{align}
where $\Delta \bm Q = \bm Q' - \bm Q$, with
\begin{align}
	\bm Q(\br, \bdu R, t) = \frac{\bm F(\br, \bdu R, t)}{m_e}+ \nabla_{\br} \cdot \bm \sigma(\br, \bdu R, t) + \bm S(\br, \bdu R, t), 
\end{align}
and $\bm Q'$ being the counterpart of the primed system. By our assumption, $\Delta \bm A(\br, \bdu R, t)$ is Taylor expandable in the neighborhood of $t=0$ with nonzero radius of convergence. Therefore we can write 
\begin{align}
	\Delta \bm A(\br, \bdu R, t) = \sum_k \Delta {\bm A}_k(\br, \bdu R) t^k. \label{Taylor}
\end{align}
Substituting \Eq{Taylor} into \Eq{DeltaA} and equating the $l$th
term of the Taylor expansion on each side, we have
\begin{align}
	&\quad \sum_{k=0}^l \frac{1}{m_e}\Big[\rho_{l-k}(\br, \bdu R)[\pt \Delta \bm A(\br, \bdu R,t)]_k- \bm J_{l-k}(\br, \bdu R)\times \Big(\nabla_{\br} \times \Delta \bm A_k(\br, \bdu R)\Big) \Big] \nonumber\\
	&\quad + [\bm Q'(\br, \bdu R,t)]_l -[\bm Q(\br, \bdu R,t)]_l =0, \label{recur}
\end{align}
where $\rho_{k}(\br, \bdu R)$ and $\bm J_{k}(\br, \bdu R)$ denote the $k$th coefficients in the Taylor expansions of $\rho$ and $\bm J$ at time $t=0$, and in general $[f(\br, \bdu R,t)]_l$ denotes the $l$th coefficient in the expansion of a function $f(\br, \bdu R, t)$ in powers of $t$ about $t=0$.

Using the relation $[\pt \Delta \bm A(\br, \bdu R,t)]_k = (k+1)\Delta \bm A_{k+1}(\br, \bdu R)$, we can rewrite \Eq{recur} as
\begin{align}
	\rho_0(\br, \bdu R) (l+1) \Delta \bm A_{l+1}(\br, \bdu R) &= -\sum_{k=0}^{l-1} \rho_{l-k}(\br, \bdu R)(k+1)\Delta \bm A_{k+1}(\br, \bdu R) \nonumber \\
	&\quad + \sum_{k=0}^{l}\bm J_{l-k}(\br, \bdu R)\times \Big(\nabla_{\br} \times \Delta \bm A_k(\br, \bdu R)\Big) \nonumber \\
	&\quad + m_e [\bm Q(\br, \bdu R,t)]_l -m_e[\bm Q'(\br, \bdu R,t)]_l. \label{recur-1}
\end{align}
On the right hand side (RHS) of \Eq{recur-1}, the first two lines explicitly depend on $\Delta \bm A_k(\br, \bdu R)$ with $k\leqslant l$. For those implicit $\Delta \bm A_k(\br, \bdu R)$s hidden in the expansion coefficients of $\bm Q'$, we show that they also depend on $\Delta \bm A_k(\br, \bdu R)$ with $k\leqslant l$ only rather than $\Delta \bm A_{l+1}(\br, \bdu R)$. 
This is because the time-dependent Schr\"odinger equation is of first order in time, which guarantees that the $l$th coefficients
in the Taylor expansion of $\Psi$ and $\Psi'$, and hence $\bm Q$ and $\bm Q'$, are completely determined by coefficients of order $k<l$ in
the Taylor expansion of $\bm A$ and $\bm A'$. Thus, \Eq{recur-1} is a recursive relation for $\Delta \bm A_{l+1}(\br, \bdu R)$. 
Using the relation
\begin{align} 
	\rho(\br, \bdu R,0)\Delta \bm A_0(\br, \bdu R) = \langle \Psi_0 | \bm {\hat j}_p(\br)  |\Psi_0 \rangle_{\bdu r} - \langle \Psi_0' | \bm {\hat j}_p(\br)  |\Psi_0' \rangle_{\bdu r},
\end{align}
we can determine the initial value of $\Delta \bm A_0(\br, \bdu R)$. Then with the recursive relation \Eq{recur-1}, we can obtain all the expansion coefficients of $\Delta \bm A(\br, \bdu R,t)$, which allows us to determine $\Delta \bm A(\br, \bdu R,t)$ within its finite radius of convergence up to some $t_c$. Then the process
can be iterated taking $t_c$ as the initial time. This completes the existence proof of our theorem. 

\section{Proof that $\ep_{\rm dyn}$ is negligible in the adiabatic limit}
The dynamical contribution to the PES is defined as
\begin{equation}
    \ep_{\rm dyn} = \langle \Phi_{\bdu R}|-i\pt |\Phi_{\bdu R}\rangle.
\end{equation}
Now we want to show that in the gauge where the vector potential is zero, if we scale the time coordinate, i.e., $\hat H(\bdu r, \bdu R, t) \rightarrow \hat {\tilde  H}(\bdu r, \bdu R, t) = \hat H(\bdu r, \bdu R, \omega t)$, then we have the following scaling relation, $\tilde \ep_{\rm dyn}(\bdu R, t) = \omega^2\ep_{\rm dyn}(\bdu R, t/\omega)$ as $\omega \rightarrow 0$. This thus implies that $\ep_{\rm dyn}$ is negligible in the adiabatic limit. To prove this, it suffices to show the same scaling relation for the corresponding dynamical force. We note
\begin{align}
    \bm F_{\rm dyn}^l &= -\nabla_{\bR_l} \ep_{\rm dyn} = -\nabla_{\bR_l} \ep_{\rm dyn} + \pt \bm A_l \nonumber \\
    &= -\nabla_{\bR_l}\langle \Phi_{\bdu R}|-i\pt |\Phi_{\bdu R}\rangle + \pt \langle \Phi_{\bdu R}|-i\nabla_{\bR_l} |\Phi_{\bdu R}\rangle \nonumber \\
    &= 2{\rm Im}\langle \pt \Phi_{\bdu R} |\nabla_l \Phi_{\bdu R}\rangle.
\end{align}
In the last line we have denoted $\nabla_l \equiv \nabla_{\bR_l}$.
The full TDSE is
\begin{align}
    i\pt \Psi(\bdu r, \bdu R, t) = -\sum_l\frac{1}{2M_l}\nabla_l^2 \Psi(\bdu r, \bdu R, t)+ \hat H_e \Psi(\bdu r, \bdu R, t).
\end{align}
Here $\hat H_e = \hat T_e + \hat V_{ee}+\hat V_{nn} + \hat V_{en}(t)$ is the electronic Hamiltonian.
Now we compute
\begin{align}
    i\pt |\chi|^2 &= i\pt \int |\Psi(\bdu r, \bdu R, t)|^2 d\bdu r \nonumber \\
    &=i\int \Psi^*(\bdu r, \bdu R, t)\pt \Psi(\bdu r, \bdu R, t) d\bdu r + i\int \Psi(\bdu r, \bdu R, t)\pt \Psi^*(\bdu r, \bdu R, t) d\bdu r \nonumber \\
    &= \int \Psi^*(\bdu r, \bdu R, t)\Big[-\sum_l\frac{1}{2M_l}\nabla_l^2 + \hat H_e \Big] \Psi(\bdu r, \bdu R, t) d\bdu r \nonumber \\
    &\quad -\int \Psi(\bdu r, \bdu R, t)\Big[-\sum_l\frac{1}{2M_l}\nabla_l^2 + \hat H_e \Big]\Psi^*(\bdu r, \bdu R, t) d\bdu r. \label{ptchi}
\end{align}
Since for each $\bdu R$, $\hat H_e$ is a Hermitian operator on the electronic Hilbert space, the $\hat H_e$ terms in \Eq{ptchi} cancel each other. Thus, \Eq{ptchi} reduces to
\begin{align}
    i\pt |\chi|^2 &= -\sum_l\frac{1}{2M_l}\int \Big[\Psi^*(\bdu r, \bdu R, t)\nabla_l^2  \Psi(\bdu r, \bdu R, t)- \Psi(\bdu r, \bdu R, t)\nabla_l^2  \Psi^*(\bdu r, \bdu R, t)\Big]d\bdu r .
\end{align}
Let $\Psi(\bdu r, \bdu R, t) = |\Psi(\bdu r, \bdu R, t)|{\rm exp}\Big\{i\gamma(\bdu r, \bdu R, t)\Big\}$, the above expression can be simplified to
\begin{align}
    \pt |\chi|^2 &= -\sum_l\frac{1}{M_l}\int \nabla_l|\Psi(\bdu r, \bdu R, t)|^2\nabla_l  \gamma(\bdu r, \bdu R, t)d\bdu r . \label{ptchi-1}
\end{align}
\Eq{ptchi-1} is reminiscent of the continuity equation in the quantum fluid dynamic (or hydrodynamic, Bohmian) representation. \cite{Bohm1951, Wyatt2005}
In the adiabatic limit, as we scale the time by $\omega$, the following relations are true:
\begin{align}
    \hat H(\bdu r, \bdu R, t) &\rightarrow \hat {\tilde H}(\bdu r, \bdu R, t) =H(\bdu r, \bdu R, \omega t), \\
    |\chi(\bdu r, \bdu R, t)| &\rightarrow |\tilde \chi(\bdu r, \bdu R, t)| = |\chi(\bdu r, \bdu R, \omega t)|, \\
    |\Psi(\bdu r, \bdu R, t)| &\rightarrow |\tilde \Psi(\bdu r, \bdu R, t)| = |\Psi(\bdu r, \bdu R, \omega t)|, \\
    \nabla_l|\Psi(\bdu r, \bdu R, t)|^2 &\rightarrow \nabla_l|\tilde \Psi(\bdu r, \bdu R, t)|^2 = \nabla_l|\Psi(\bdu r, \bdu R, \omega t)|^2.
\end{align}
It follows that $\pt |\tilde \chi|^2\Big|_{t = T/\omega} = \omega \pt |\chi|^2\Big|_{t = T}$ for any given $T$. Thus combining with \Eq{ptchi-1}, we have the following scaling relation for $\gamma$,
\begin{align}
    \sum_l\frac{1}{M_l}\int \nabla_l|\Psi(\bdu r, \bdu R, T)|^2 \Big[\nabla_l \tilde \gamma(\bdu r, \bdu R, T/\omega)- \omega \nabla_l \gamma(\bdu r, \bdu R, T)\Big] d\bdu r=0.
\end{align}
This is true for all possible $\bdu R$, which implies
\begin{align}
    \nabla_l \tilde \gamma(\bdu r, \bdu R, T/\omega)- \omega \nabla_l \gamma(\bdu r, \bdu R, T) = 0. \label{rl1}
\end{align}
Similarly, by computing $\pt \int|\Psi(\bdu r, \bdu R, t)|^2 d\bdu R$, one can derive a similar equation to \Eq{ptchi-1} as the following,
\begin{align}
    \pt \int|\Psi(\bdu r, \bdu R, t)|^2 d\bdu R &= -N_e\frac{1}{m_e}\int \nabla_\br|\Psi(\bdu r, \bdu R, t)|^2\nabla_\br  \gamma(\bdu r, \bdu R, t)d\bdu R . \label{ptchi-2}
\end{align}
Here $\bdu r = (\br,\br_2,\cdots, \br_{N_e})$. By carrying out the same analysis, one can derive the analogous equation to \Eq{rl1} as
\begin{align}
    \nabla_\br \tilde \gamma(\bdu r, \bdu R, T/\omega)- \omega \nabla_\br \gamma(\bdu r, \bdu R, T) = 0. \label{rl2}
\end{align}
\Eq{rl1} and \Eq{rl2} thus suggest that
\begin{equation}
    \tilde \gamma(\bdu r, \bdu R, T/\omega) = \omega \gamma(\bdu r, \bdu R, T) + G(T),
\end{equation}
where $G(T)$ is some constant function of $T$. Replacing $T$ by $\omega t$ and taking the time derivative, we have
\begin{equation}
    \pt \tilde \gamma(\bdu r, \bdu R, t) = \omega^2 \gamma(\bdu r, \bdu R, \omega t) + \omega G(\omega t). \label{rl3}
\end{equation}
Now we express $\bm F_{\rm dyn}^l$ in terms of $|\Psi|$, $|\chi|$ and $\gamma$. Since the dynamical force is a gauge-invariant quantity, we can evaluate it under the gauge of $\chi = |\chi|$ so that $\Phi_{\bdu R} = |\frac{\Psi}{\chi}|e^{i\gamma}$.
\begin{align}
    \ep_{\rm dyn}(\bdu R,t) = \langle \Phi_{\bdu R} | -i\pt |\Phi_{\bdu R} \rangle = {\rm Im}\langle \Phi_{\bdu R} | \pt |\Phi_{\bdu R} \rangle = \frac{1}{|\chi(\bdu R,t)|^2}\int |\Psi(\bdu r, \bdu R, t)|^2\pt \gamma(\bdu r, \bdu R, t) d\bdu r.
\end{align}
\begin{align}
    \bm A_l(\bdu R, t) = \langle \Phi_{\bdu R} | -i\nabla_l |\Phi_{\bdu R} \rangle =\frac{1}{|\chi(\bdu R,t)|^2}\int |\Psi(\bdu r, \bdu R, t)|^2\nabla_l \gamma(\bdu r, \bdu R, t) d\bdu r.
\end{align}
Therefore,
\begin{align}
    \bm F_{\rm dyn}^l(\bdu R,t) &= \pt \bm A_l(\bdu R, t) - \nabla_l \ep_{\rm dyn}(\bdu R,t) \nonumber \\
    &= \int \Big\{\pt \Big[\frac{|\Psi(\bdu r, \bdu R, t)|^2}{|\chi(\bdu R,t)|^2}\nabla_l \gamma(\bdu r, \bdu R, t)\Big] - \nabla_l \Big[\frac{|\Psi(\bdu r, \bdu R, t)|^2}{|\chi(\bdu R,t)|^2}\pt \gamma(\bdu r, \bdu R, t)\Big] \Big\}d\bdu r \nonumber \\
    &= \int \Big\{\pt \frac{|\Psi(\bdu r, \bdu R, t)|^2}{|\chi(\bdu R,t)|^2}\nabla_l \gamma(\bdu r, \bdu R, t) - \nabla_l \frac{|\Psi(\bdu r, \bdu R, t)|^2}{|\chi(\bdu R,t)|^2}\pt \gamma(\bdu r, \bdu R, t) \Big\}d\bdu r.
\end{align}
Now if we scale the time, then the first term in the integral will factor out $\omega^2$ (the time derivative and the gradient will each factor out an $\omega$); the second term becomes (by \Eq{rl3})
\begin{align}
    -\nabla_l \frac{|\tilde \Psi(\bdu r, \bdu R, t)|^2}{|\tilde \chi(\bdu R,t)|^2} \pt \tilde\gamma(\bdu r, \bdu R, t) = -\nabla_l \frac{|\Psi(\bdu r, \bdu R, \omega t)|^2}{|\chi(\bdu R, \omega t)|^2} \Big[ \omega^2 \pt \gamma(\bdu r, \bdu R, \omega t)+\omega G(\omega t)\Big]. \label{xx}
\end{align}
Once taking the integration over $\bdu r$, the second term in the square bracket of \Eq{xx} vanishes, leaving only the $\omega^2$ term. Thus
\begin{equation}
    \tilde {\bm F}_{\rm dyn}^l(\bdu R,t) = \omega^2 \bm F_{\rm dyn}^l(\bdu R,\omega t), \label{Fdyn-scale}
\end{equation}
and completes our proof.

\section{Some details of our driven-proton-transfer model}
The full time dependent electron-nuclear Hamiltonian for our two-site Hubbard model is
\begin{equation}
    \bm {\hat H}(R,t) = -\frac{1}{2M}\nabla^2+\bm H_e(R)+\bm V(R,t).
\end{equation}
In the basis representation of the three singlet states, namely, $\varphi_1 = |1_\ua 1_\da\rangle$, $\varphi_2 = \frac{1}{\sqrt 2}(|1_\ua 2_\da\rangle-|1_\da 2_\ua\rangle$), and $\varphi_3 = |2_\ua 2_\da\rangle$, the intrinsic electronic Hamiltonian can be described as
\begin{equation}
    \bm H_e(R) = \begin{pmatrix}
    U_1 + \Delta \epsilon(R), & -\sqrt 2 \tau(R), & 0 \\
    -\sqrt 2\tau(R), & 0, & -\sqrt 2\tau(R) \\
    0, & -\sqrt 2\tau(R), & U_2 - \Delta \epsilon(R)
    \end{pmatrix}.
\end{equation}
The external driving potential is
\begin{equation}
    \bm V(R,t) = w(R)\hat n \cos\omega t ,
\end{equation}
with $\hat n = \rm{diag}(-1,0,1)$ is defined as the density operator in this model. The spatial functions in $\bm H_e$ and $\bm V(R,t)$ are
\begin{equation}
    \tau(R) = t_0\cos^2 \frac{2\pi R}{L}+t_1,
\end{equation}
\begin{equation}
    \Delta \epsilon (R) = \Delta I + \gamma \sin^2 \frac{2\pi R}{L},
\end{equation}
\begin{equation}
    w(R) = V_0\sin(\frac{2\pi R}{L}).
\end{equation}
And the model parameters are listed in Table~\ref{para}.
\setlength{\tabcolsep}{12pt}
\setlength{\extrarowheight}{6pt}
\begin{table}[H]
\caption{Model parameters. All energies are in unit of eV. \label{para}}
\centering
\begin{tabular}{ccccccccc}
\hline
 $M (m_e)$ & $L$ (Bohr) & $U_1 $ & $U_2$ & $t_0$ & $t_1$ & $V_0$ & $\gamma$ & $\Delta I$ \tabularnewline
\hline
2000 & 3.0 & 0.08 & 0.13 & 0.005 & 0.005 & 0.08 & 0.1 & 0.05 \tabularnewline
\hline
\end{tabular}
\end{table}

\begin{figure}[H]
\includegraphics[scale=0.6]{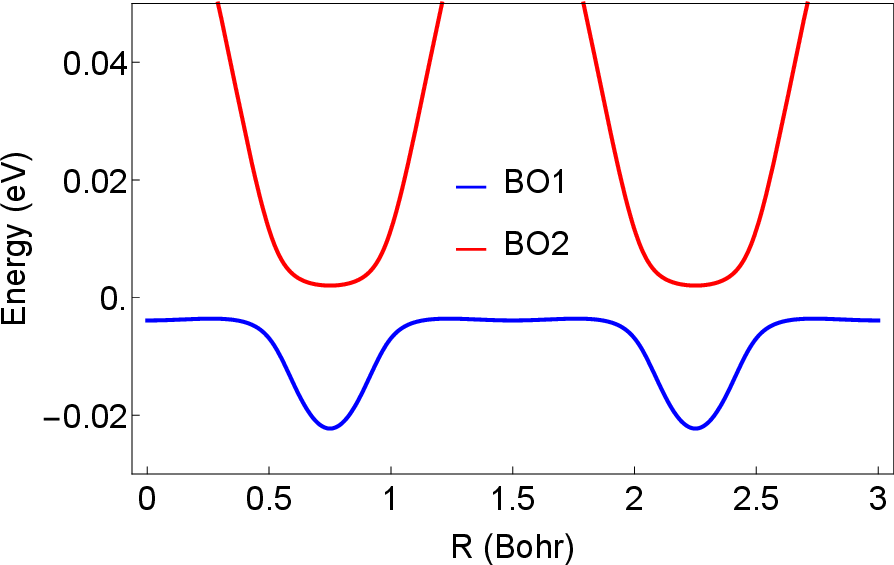}
\centering
\caption{
The lowest two BO PESs of $\bm H_e(R)$.  \label{PESBO}}
\end{figure}

The lowest two BO PESs of $\bm H_e(R)$ are shown in Fig~\ref{PESBO}. As can be seen, the ground BO surface is a double well potential, with avoided crossings with the first excited state surface at the shoulders of the wells.

\begin{figure}[H]
\includegraphics[scale=0.53]{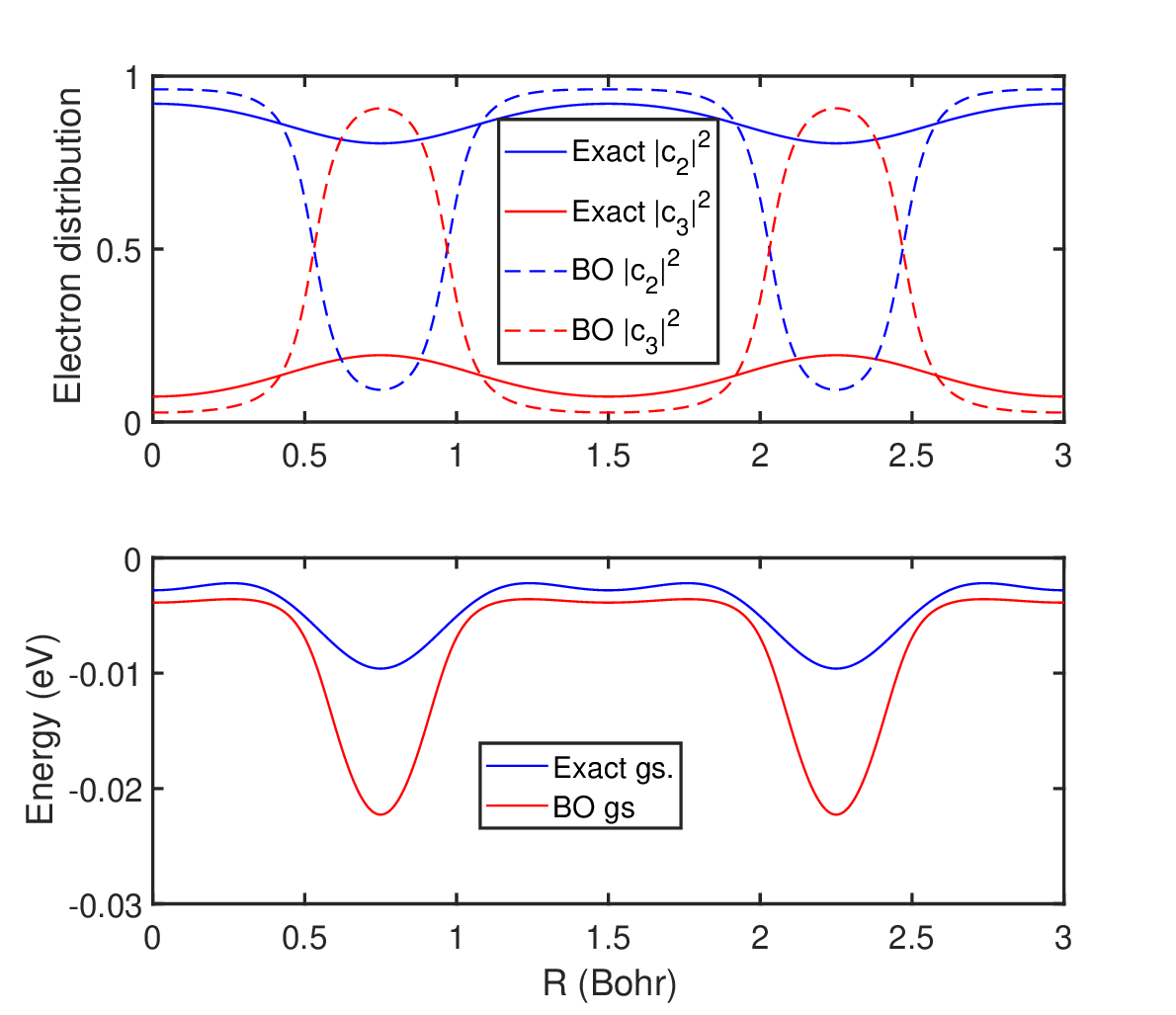}
\centering
\caption{
Upper panel: populations of the many-body configurations in the conditional
electronic wave functions $|\bm \Phi_R\rangle$ and $|\bm \Phi_R^{\rm BO}\rangle$ at time $t=0$ (we assume the total electron-nuclear wave function is at its instantaneous ground state); a third higher-energy state $\varphi_1$
has negligible population $|c_1|^2$ for all $R$ and is not shown.  Lower panel: comparison between the
exact and BO ground state potential energy surfaces in our model. Here we have switched off the external potential $\bm V$. \label{gsPES0}}
\end{figure}

\begin{figure}[H]
\includegraphics[scale=0.53]{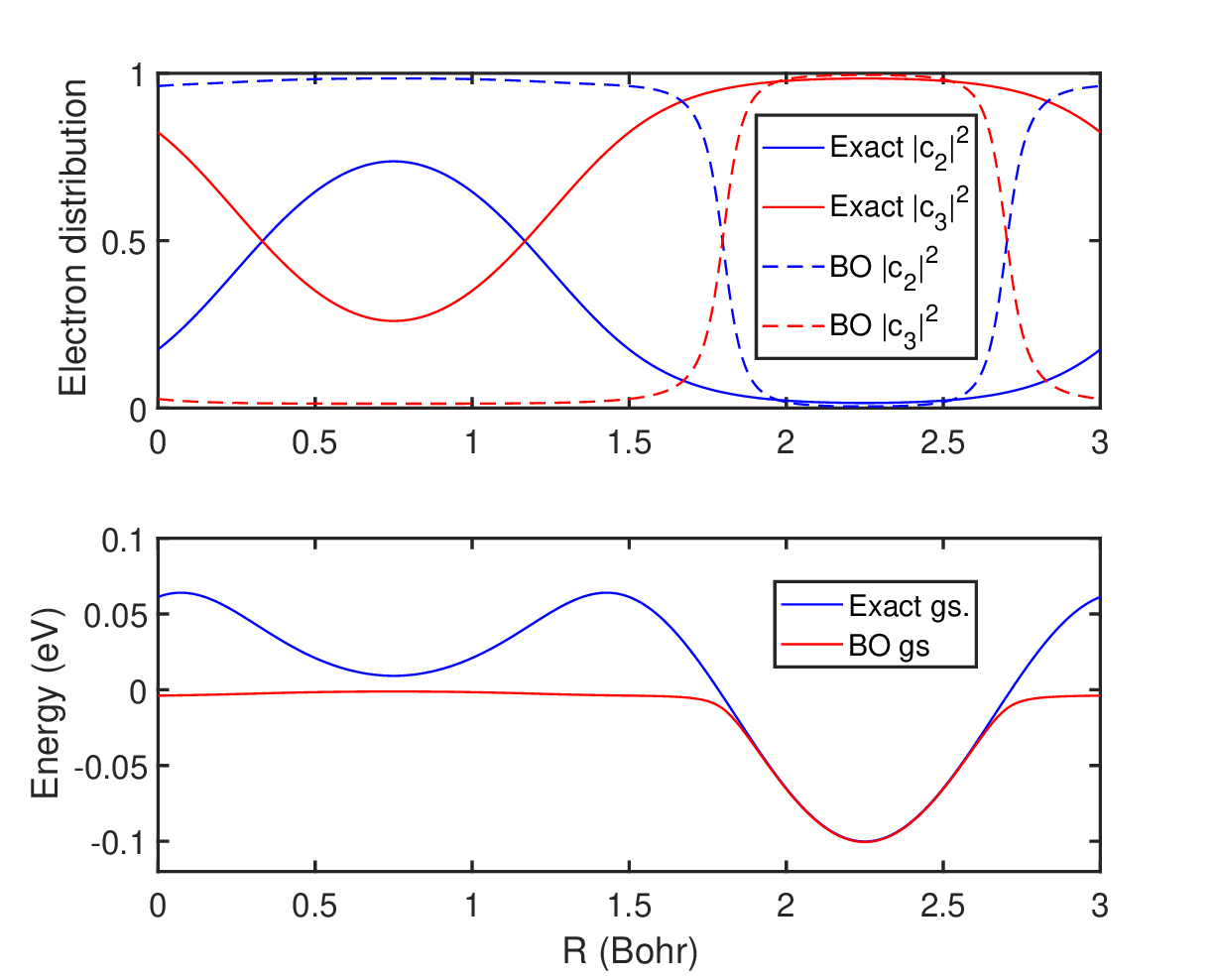}
\centering
\caption{
Same variables as in Fig~\ref{gsPES0}. The distinction is that we have switched on the external potential $\bm V$.  \label{gsPES}}
\end{figure}

\begin{figure}[H]
\includegraphics[scale=0.6]{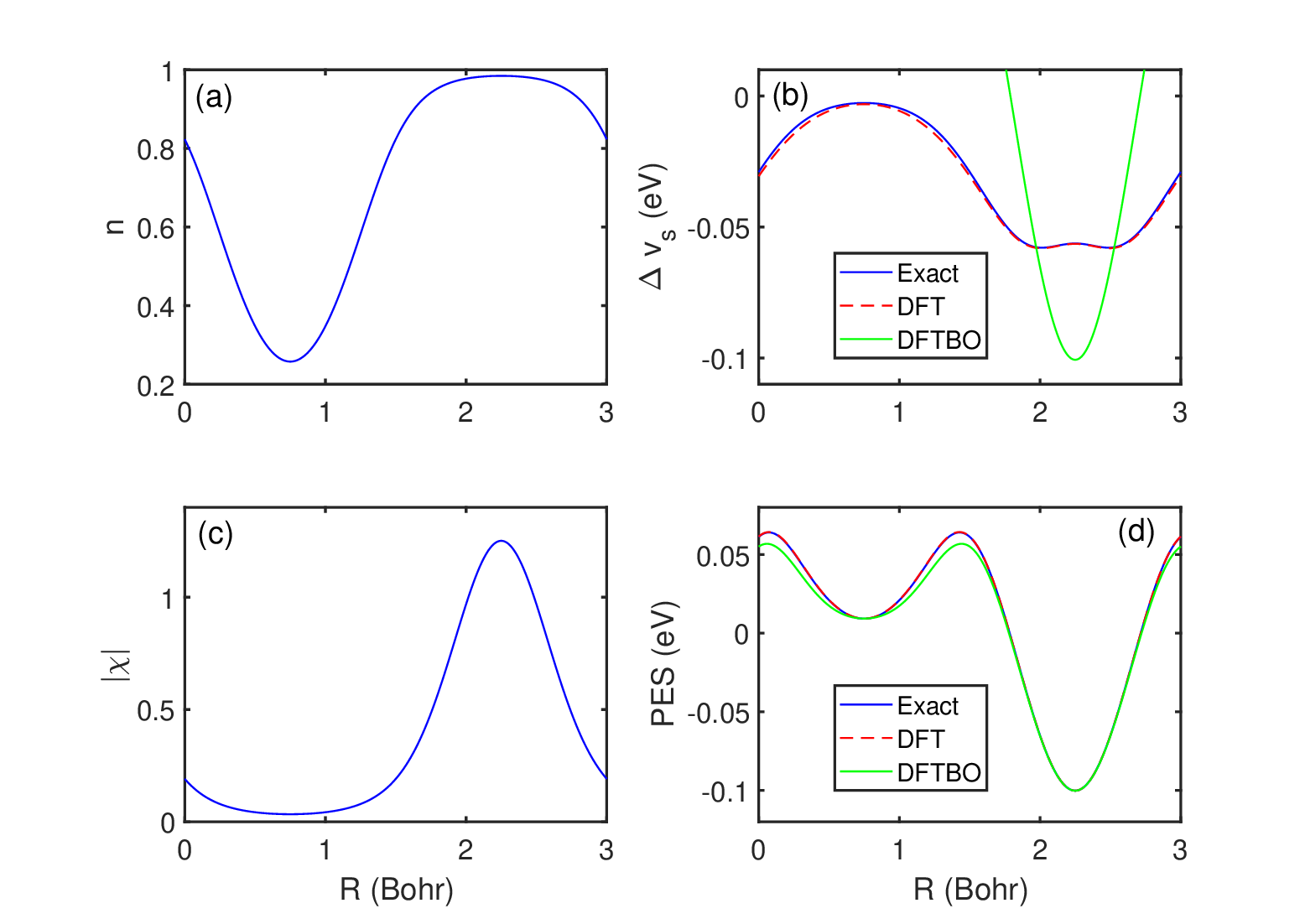}
\centering
\caption{
(a) initial conditional electronic density; (b) comparing the static Kohn-Sham potential between the exact, the DFT and the DFTBO (without geometric correction); (c) initial $|\chi|$; (d) comparing initial PES between the exact, the DFT and the DFTBO (without geometric correction).   \label{gsDFT}}
\end{figure}

The ground state BO electronic distributions and PES with and without the external potential $\bm V(R,0)$ are shown in Figs~\ref{gsPES0} and \ref{gsPES}. Comparison has been made between the BO and the exact ground state. The initial ground state density $n(R)$ and $|\chi(R)|$ and the corresponding KS potential and PES are shown in Fig~\ref{gsDFT}.

\begin{figure}[H]
\includegraphics[scale=0.6]{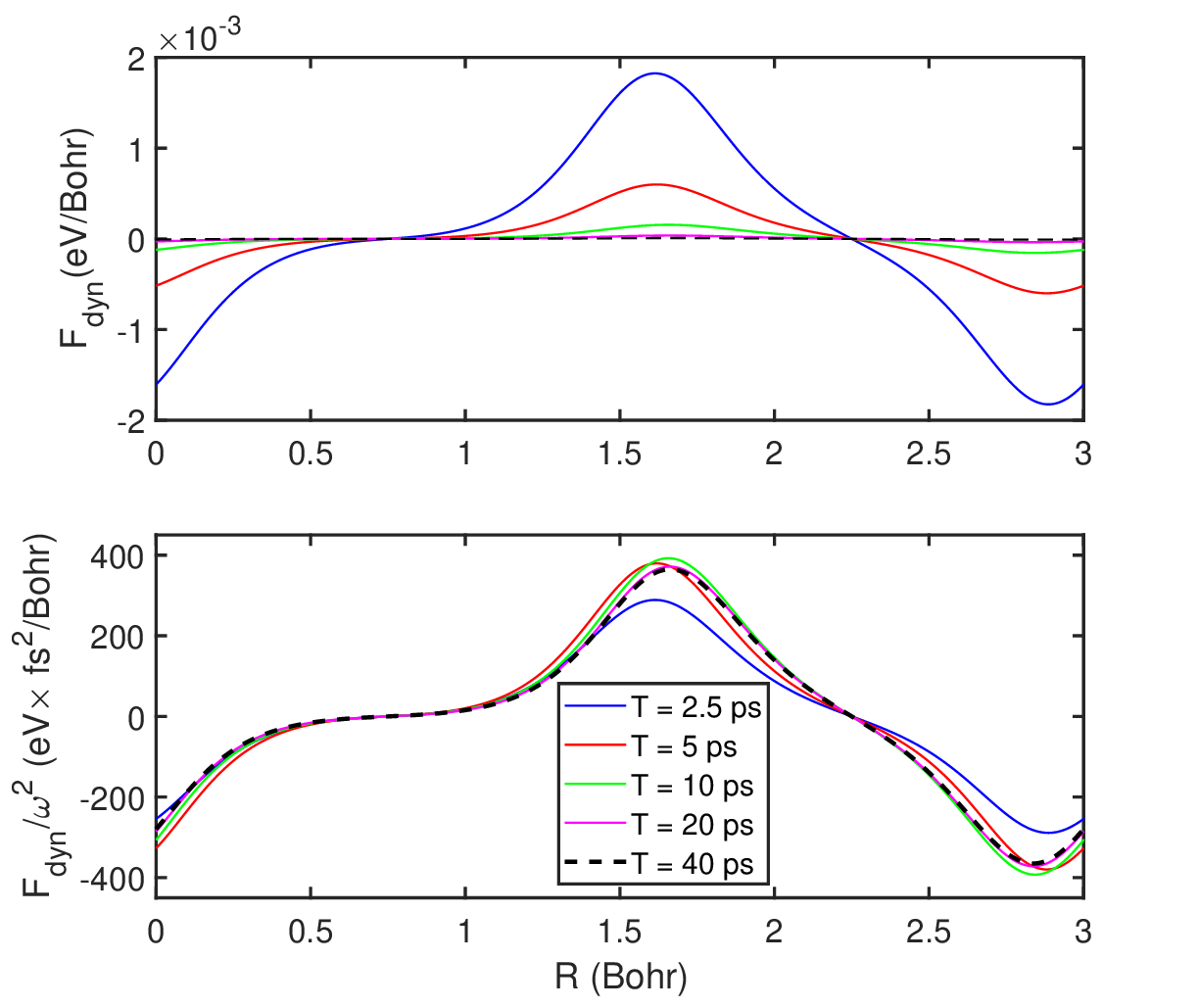}
\centering
\caption{
Comparing (a) $F_{\rm dyn}$ and (b) $\frac{F_{\rm dyn}}{\omega^2}$ at $t=\frac{T}{6}$ for different choice of $T$.   \label{Fdyn}}
\end{figure}

In Fig~\ref{Fdyn}, we numerically verify \Eq{Fdyn-scale} by showing $F_{\rm dyn}$ and $\frac{F_{\rm dyn}}{\omega^2}$ at $t=\frac{T}{6}$ for different choice of $T$. As can be seen, $\frac{F_{\rm dyn}}{\omega^2}$ converges as $T\rightarrow \infty$, which implies \Eq{Fdyn-scale} is true.

\begin{figure}[H]
\includegraphics[scale=0.6]{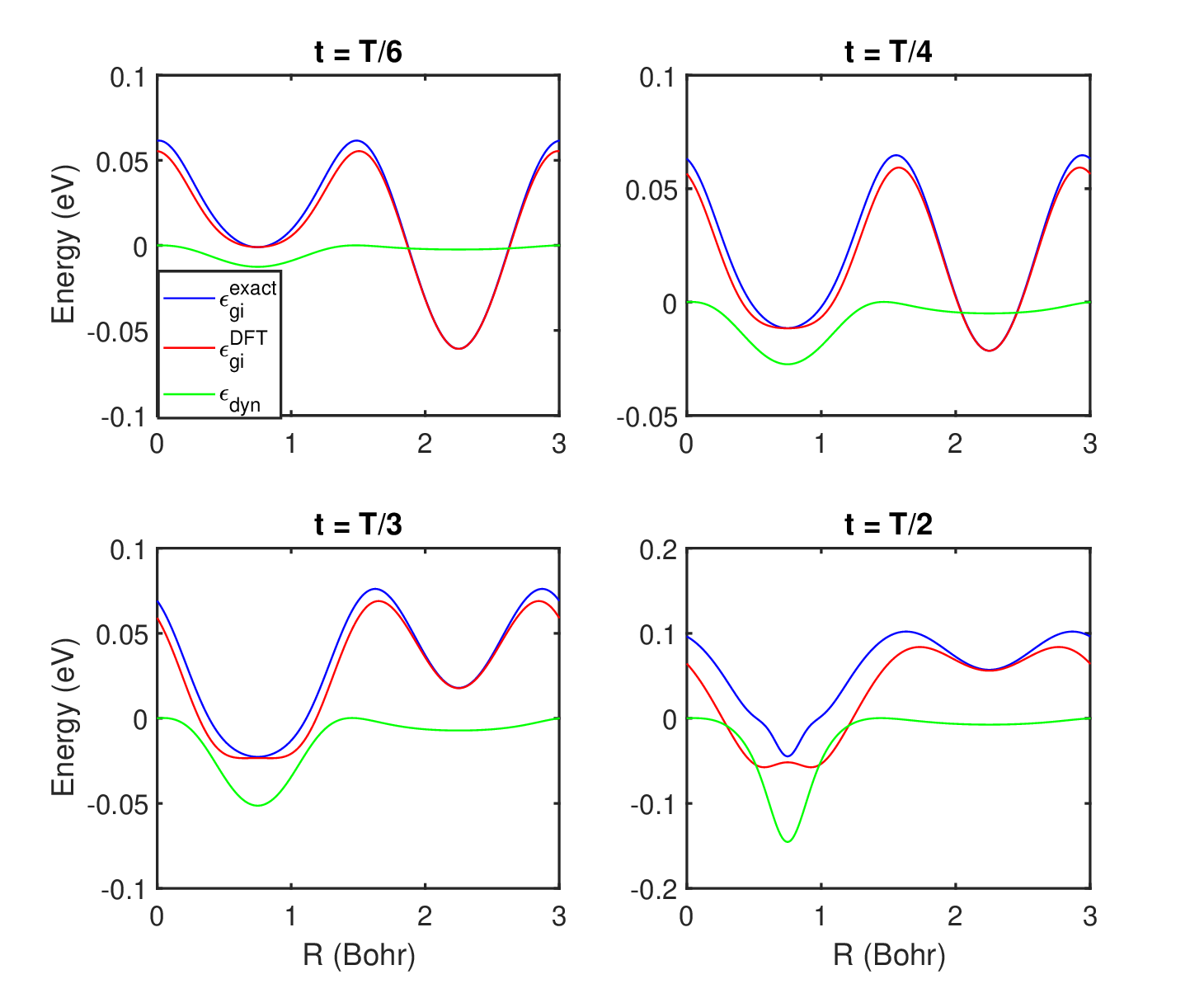}
\centering
\caption{
Comparing, during time evolution in the diabatic regime, $\ep_{\rm dyn}$ with the gauge independent part of the PES from the exact ($\epsilon_{\rm gi}^{\rm exact} = \langle \bm \Phi_R |\bm H_e | \bm \Phi_R\rangle + \frac{1}{2M}\langle \nabla \bm \Phi_R |\nabla \bm \Phi_R\rangle$) and DFT calculation ($\epsilon_{\rm gi}^{\rm DFT} = \epsilon_{\rm BO}^{\rm approx}[n_R]+\epsilon_{\rm geo}[n_R,\nabla n_R]$). Here we have chosen $T=0.02$ ps and set $\ep_{\rm dyn}(R=0)=0$. The importance of $\ep_{\rm dyn}$ in the diabatic regime is highlighted.    \label{Edyn}}
\end{figure}

As a remark, $F_{\rm dyn}$ (and also $\epsilon_{\rm dyn}$) is not negligible in the diabatic or intermediate regimes. This is demonstrated in Fig~\ref{Edyn}, where we compare the energy scale of $\epsilon_{\rm dyn}$ with the gauge independent part of the PES, $\epsilon_{\rm gi}^{\rm exact} = \langle \bm \Phi_R |\bm H_e | \bm \Phi_R\rangle + \frac{1}{2M}\langle \nabla \bm \Phi_R |\nabla \bm \Phi_R\rangle$, in the diabatic regime. As shown, $\epsilon_{\rm dyn}$ plays a nontrivial role and is on the similar magnitude as $\epsilon_{\rm gi}$. Dropping this term in the TDDFT time evolution would definitely lead to a failure even if the $\epsilon_{\rm gi}$ part of the PES is decently approximated. In the intermediate regimes, $\epsilon_{\rm dyn}$ also has some non-negligible effect (not shown here). Thus, approximating $\epsilon_{\rm dyn}$ as a functional of $n_R$ and $|\chi|$ in the most general case is a necessary step in the TDDFT functional development for electron-nuclear systems.

\section{Adiabatic extension approximation for our model}

In the absence of the external time dependent potential, we use the functional developed in \cite{Li18084110} to describe the ground state, which is given by
\begin{align}
    \ep[n_R] = \ep_{\rm BO}[n_R] + \ep_{\rm geo}[n_R,\nabla n_R]. \label{ep}
\end{align}
Here $\ep_{\rm BO}[n_R]$ is the exact BO ground state functional, which for our two-site Hubbard model, can be defined through the following constrained search, \cite{Li18084110}
\begin{align}
    \ep_{\rm BO}[n] &= \min_{0\leqslant u \leqslant \sqrt{1-|n|}} \Big\{-2\sqrt 2 \tau\sqrt{(1-\frac{|n|}{1-u^2})\frac{|n|}{1-u^2}\Big[1+ u\sqrt{2-u^2}\;\Big]}\nonumber \\
    &\quad + \frac{|n|}{2(1-u^2)}(\tilde U_1+\tilde U_2) \Big\}+ \frac{1}{2}n(\tilde U_2-\tilde U_1), \qquad (n\neq 0), \label{exact_func}
\end{align}
and
\begin{equation}
    \ep_{\rm BO}[0] = -\sqrt{4\tau^2+\frac{1}{16}(\tilde U_1+\tilde U_2)^2}+ \frac{1}{4}(\tilde U_1+\tilde U_2).
\end{equation}
Here $\tilde U_1 = U_1 + \Delta \ep(R)$ and $\tilde U_2 = U_2 - \Delta \ep(R)$. $\ep_{\rm BO}$ has parametric dependence on $R$ but this is suppressed here and below for the ease of notations. In practice, assuming the density is positive, we approximate the exact BO functional by the following form, \cite{Li18084110}
\begin{align}
    \ep_{\rm BO}^{\rm approx}[n] = -2\sqrt 2 \tau \sqrt{n(1-n)} + n \tilde U_2.
\end{align}
In the time dependent problem, we can add the external potential term onto $\tilde U_2(R)$, which leads to the following form for the instantaneous BO ground state energy,
\begin{align}
    \ep_{\rm BO}^{\rm approx}[n_R(t)] = -2\sqrt 2 \tau(R) \sqrt{n_R(t)[1-n_R(t)]} + n_R(t)[ \tilde U_2(R) + w(R) \cos\omega t] .
\end{align}
On the other hand, the geometric correction in \Eq{ep} under the local conditional density approximation (LCDA) \cite{Li18084110} reads
\begin{align}
    \ep_{\rm geo}[n_R,\nabla n_R] = \frac{1}{2M}f(n_R)(\nabla n_R)^2,
\end{align}
where
\begin{equation}
    f(n_R) = \frac{1}{4n_R(1-n_R)}.
\end{equation}
The Kohn-Sham system for our two-site Hubbard model is given by the folloing $2\times 2$ matrix,
\begin{equation}
    \hat h_s = \begin{pmatrix}
        -\frac{1}{2}\Delta v_s & -\tau\\
        -\tau & \frac{1}{2}\Delta v_s
    \end{pmatrix},
\end{equation}
where the Kohn-Sham potential can also be written as the sum of a BO component and a geometric correction,
\begin{equation}
    \Delta v_s = \Delta v_s^{\rm BO} + v_{\rm geo}.
\end{equation}
Here
\begin{equation}
    \Delta v_s^{\rm BO} = \frac{\partial \ep_{\rm BO}}{\partial n}-\frac{\partial T_s}{\partial n},
\end{equation}
where $T_s=-2\tau\sqrt{1-n^2}$ is the non-interacting kinetic energy. Then within our approximation the Kohn-Sham potential is given by
\begin{align}
    \Delta v_s^{\rm BO, approx} &= \frac{\partial \ep_{\rm BO}^{\rm approx}}{\partial n} -\frac{\partial T_s}{\partial n} \nonumber \\
    &= -\sqrt 2 \tau\frac{1-2n_R}{\sqrt{n_R(1-n_R)}}+\tilde U_2(R) + w(R) \cos\omega t - \frac{2n\tau}{\sqrt{1-n_R^2}}.
\end{align}
The geometric correction to the Kohn-Sham potential is not a simple functional derivative of $\ep_{\rm geo}$ with respect to $n_R$, but also has dependence on $|\chi|^2$, \cite{Li18084110}
\begin{align}
    v_{\rm geo} = -\frac{1}{M}\Big[\frac{1}{2}f'(n_R)(\nabla n_R)^2 + f(n_R) \nabla^2 n_R + f(n_R) \nabla \ln|\chi(R)|^2 \nabla n_R \Big].
\end{align}
Here $f'$ is the derivative of the function $f$.
In the adiabatic extension approximation, we substitute the instantaneous time-evolving $n_R(t)$ and $|\chi(R,t)|^2$ into the functional form of $\ep$ and $\Delta v_s$ and evaluate the corresponding PES and KS potential, then perform the time evolution of the following coupled equations,
\begin{align}
    i\pt \begin{pmatrix}
        c_1(R,t)\\
        c_2(R,t)
    \end{pmatrix}
    &= \begin{pmatrix}
        -\frac{1}{2}\Delta v_s(R,t) & -\tau(R)\\
        -\tau(R) & \frac{1}{2}\Delta v_s(R,t)
    \end{pmatrix}
    \begin{pmatrix}
        c_1(R,t)\\
        c_2(R,t)
    \end{pmatrix}, \label{TDKS} \\
    i\pt \chi(R,t) &= -\frac{1}{2M}\nabla^2 \chi(R,t) + \ep(R,t)\chi(R,t).
\end{align}
Here $n_R(t) = |c_2(R,t)|^2-|c_1(R,t)|^2$ and $|c_1(R,t)|^2+|c_2(R,t)|^2=1$.

\section{Details of computing the exact PES and Kohn-Sham potential}

\subsection{Exact solution of TDSE}
The exact PES and the Kohn-Sham potential can be derived from the exact time dependent $\bm \Psi$. First of all, the full time dependent Schr{\"o}dinger equation for our model is
\begin{equation}
    i\pt \bm \Psi(R,t) = \bm {\hat H}(R,t)\bm \Psi(R,t). \label{TDSE}
\end{equation}
To solve this equation numerically, we expand $\bm \Psi$ as
\begin{equation}
    \bm \Psi(R,t) = \sum_{nk}C_{nk}(t)B_n(R)\hat e_k.
\end{equation}
Here, we adopt the trigonometric functions $(1, \sin\frac{2\pi kR}{L}, \cos\frac{2\pi kR}{L}, k=\pm1, \pm2, \cdots)$ as real space basis functions $B_n(R)$.
$\hat e_k$ are the three-component electronic basis vectors, corresponding to the electronic states $\varphi_k$; the $k$th component of $\hat e_k$ is 1 and the rest are 0. This transforms \Eq{TDSE} into an ordinary differential equation. Then with a given initial condition for $\bm \Psi(R,0)$, 
we can perform numerical integration and obtain the exact solution of $\bm \Psi(R,t)$ up to a specified numerical accuracy.

With $\bm \Psi(R,t)$, we then compute the exact $n_R(t)$ and $|\chi(R,t)|$. 
The exact PES and Kohn-Sham potential can be reverse engineered as follows.

\subsection{Exact PES}
In the most general gauge choice where the vector potential is nonzero, the nuclear wave function satisfies the following time dependent equation,
\begin{equation}
    i\pt \chi = \frac{1}{2M}(-i\nabla+A)^2 \chi + \epsilon\chi.
\end{equation}
Now under the gauge of $\chi=|\chi|$, denote the scalar and vector potential as $\ep_1$ and $A_1$. Collecting the real part of the nuclear equation, we have
\begin{equation}
    \frac{1}{2M}(-\nabla^2+A_1^2) |\chi| + \epsilon_1|\chi| = 0,
\end{equation}
so that
\begin{equation}
    \epsilon_1 = \frac{1}{2M}\Big(\frac{\nabla^2 |\chi|}{|\chi|} - A_1^2\Big).
\end{equation}
Now denote the exact PES under the gauge of $A=0$ as $\ep_{\rm exact}$. Using the fact that $\nabla \epsilon - \pt A$ is a gauge invariant quantity, we arrive at
\begin{equation}
    \nabla \ep_{\rm exact} = \nabla \epsilon_1 - \pt A_1 =  \frac{1}{2M}\nabla \Big(\frac{\nabla^2 |\chi|}{|\chi|} - A_1^2\Big)- \pt A_1.
\end{equation}
The exact PES is determined up to an additive constant function of $t$. In this work, we set $\ep_{\rm exact}(R=0, t) = \ep_{\rm DFT}(R=0,t)$, then the exact PES can be calculated as
\begin{align}
    \ep_{\rm exact}(R, t) = \ep_{\rm DFT}(0,t) + \int_0^R \nabla \ep_{\rm exact}(R',t)dR'.
\end{align}

\subsection{Exact Kohn-Sham potential}
In \Eq{TDKS}, since $|c_1|^2 + |c_2|^2=1$ and $|c_2|^2-|c_1|^2 =n$, we can solve for $|c_1|$ and $|c_2|$ as
\begin{align}
    |c_1| &= \sqrt{\frac{1-n}{2}}, \\
    |c_2| &= \sqrt{\frac{1+n}{2}}.
\end{align}
Now let $c_1 = |c_1|e^{is_1}$ and $c_2 = |c_2| e^{i(s_1+\Delta s)}$. By the equation of motion,
\begin{align}
    i\pt n = \langle \bm \Phi_{\rm KS}|[\hat n, \hat h_s]|\bm \Phi_{\rm KS}\rangle. \label{EOM-KS}
\end{align}
Here $\Phi_{\rm KS} = (c_1, c_2)^T$ and $\hat n = {\rm diag}(-1,1)$, substituting them into \Eq{EOM-KS}, we arrive at
\begin{align}
    i\pt n = 2\tau (c_1^* c_2 - c_2^* c_1) = 4i\tau |c_1||c_2|\sin \Delta s.
\end{align}
Therefore,
\begin{equation}
    \sin \Delta s = \frac{\pt n}{4\tau |c_1||c_2|} = \frac{\pt n}{2\tau \sqrt{1-n^2}}. \label{dsin}
\end{equation}
Now we rewrite \Eq{TDKS} as
\begin{align}
    i\pt c_1 &= -\frac{1}{2}\Delta v_s c_1 - \tau c_2, \\
    i\pt c_2 &= -\tau c_1 + \frac{1}{2}\Delta v_s c_2.
\end{align}
by which we can solve for $\Delta v_s$ as
\begin{align}
    \Delta v_s &= - 2 i\pt \ln c_1 -2\tau \frac{c_2}{c_1}, \label{vs1} \\
    \Delta v_s &= 2i\pt \ln c_2 + 2\tau \frac{c_1}{c_2}. \label{vs2}
\end{align}
\Eq{vs1}+\Eq{vs2} leads to
\begin{align}
    \Delta v_s &= i\pt \ln \frac{c_2}{c_1} + \tau (\frac{c_1}{c_2}-\frac{c_2}{c_1}),
\end{align}
which after simplification reads
\begin{align}
    \Delta v_s &= -\pt \Delta s -\tau \cos \Delta s \frac{2n}{\sqrt{1-n^2}}  \\
            &= -\frac{\pt \sin\Delta s}{\cos\Delta s} -\tau \cos \Delta s \frac{2n}{\sqrt{1-n^2}}.
\end{align}
Here $\cos\Delta s = \pm \sqrt{1-\sin^2\Delta s}$ can take positive or negative root. However, in the adiabatic limit, $\pt n = 0$ so that $\sin\Delta s=0$ and $\Delta v_s$ should reduce to the correct static KS potential, which is given by $-\frac{2n\tau}{\sqrt{1-n^2}}$ \cite{Li18084110}. This suggests that $\cos\Delta s$ should take the positive root. Therefore,
\begin{align}
    \Delta v_s  &= -\frac{\pt \sin\Delta s}{\sqrt{1-\sin^2\Delta s}} -\tau \sqrt{1-\sin^2\Delta s} \frac{2n}{\sqrt{1-n^2}}. \label{dvs}
\end{align}
A few remarks on the non-interacting $v$-representability are as follows. In the main text, we have assumed that the time evolving conditional electronic density is non-interacting $v$-representable. For our model, the validity of this assumption is completely determined by the right hand side (RHS) of \Eq{dsin}. In particular, if $|\frac{\pt n}{2\tau \sqrt{1-n^2}}|$ is bounded by 1, then $\Delta s$ is well-defined and by \Eq{dvs} we can determine the time evolving Kohn-Sham potential; this suggests that the non-interacting $v$-representability assumption is valid. If $|\frac{\pt n}{2\tau \sqrt{1-n^2}}|>1$ for some $R$, then the above assumption is no longer valid. In the adiabatic limit, however, since $\pt n \rightarrow 0$, $|\frac{\pt n}{2\tau \sqrt{1-n^2}}| \rightarrow 0$. Thus the non-interacting $v$-representability assumption has to be true.

\begin{figure}[H]
\includegraphics[scale=0.5]{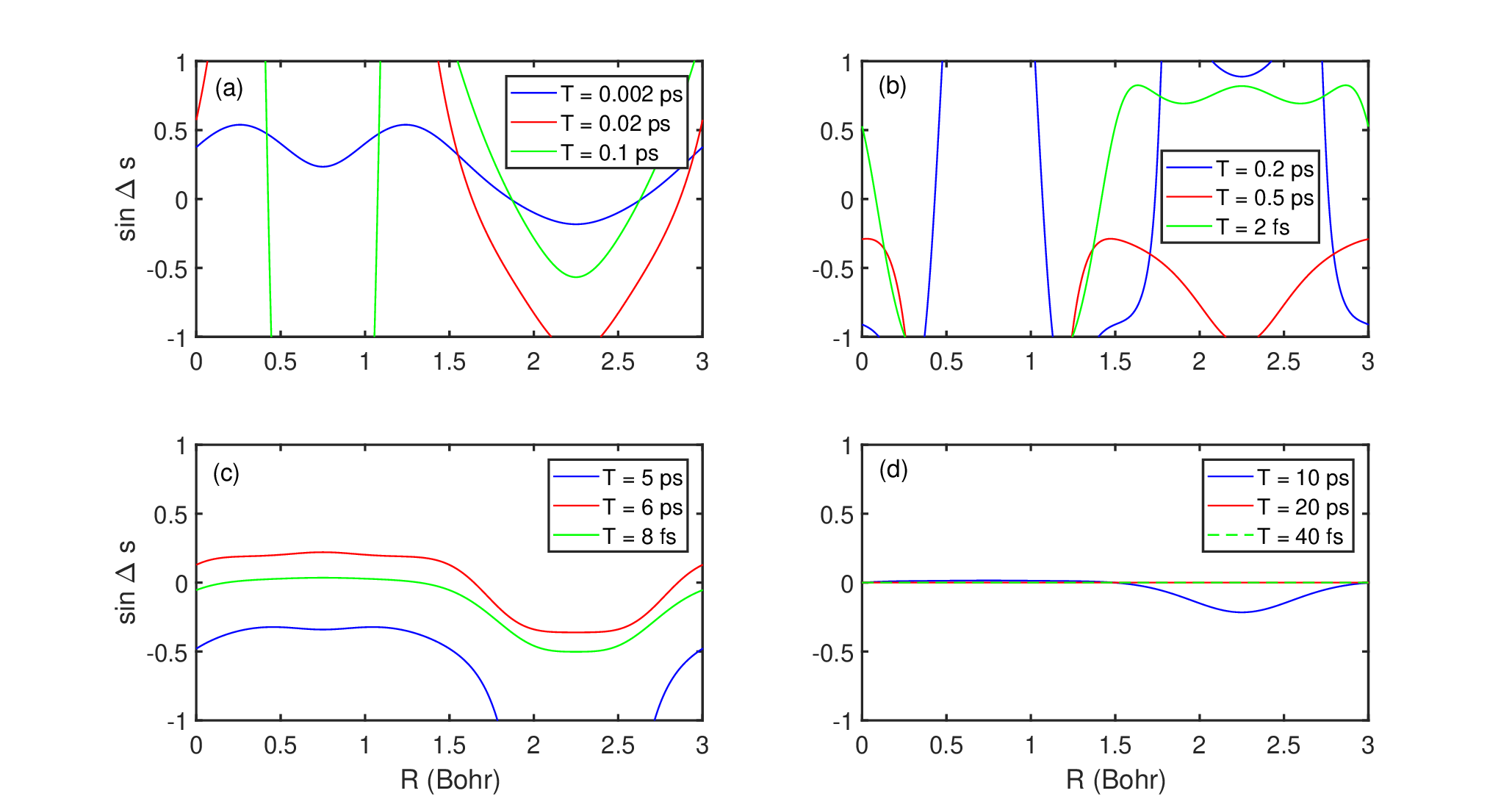}
\centering
\caption{
$\sin \Delta s$ for different regimes (from diabatic $T\rightarrow 0$ to adiabatic $T\rightarrow \infty$) evaluated at $t=\frac{T}{2}$. \label{sins}}
\end{figure}

Shown in Fig~\ref{sins} are the RHS of \Eq{dsin} evaluated at $T=\frac{T}{2}$ for different regimes. From the figure, we can deduce that for the diabatic regime ($T\rightarrow 0$) or the adiabatic regime ($T\rightarrow \infty$), the time evolving densities are non-interacting $v$-representable; while for the intermediate regimes, the corresponding densities are not.

Although the non-$v$-representable densities could be an artifact of the truncated dimensionality of the electronic Hilbert space- for continuous electron densities such scenario may or may not occur, here our example should give a warning to the $v$-representable assumption.
